\documentclass[11pt]{article}
\usepackage[utf8]{inputenc}
\usepackage[a4paper]{geometry}
\usepackage{graphicx}
\usepackage{amsmath}
\usepackage{authblk}
\usepackage{hyperref}
\usepackage{float}
\usepackage[table]{xcolor}
\usepackage{ulem}
\usepackage{longtable}
\usepackage{multirow}
\usepackage{color}
\usepackage[compact]{titlesec}

\usepackage{graphicx}
\graphicspath{ {./images/} }

\titlespacing*{\subsubsection}
  {0pt}{0.5\baselineskip}{0\baselineskip}

\title{ The distribution of loss to future USS pensions due to the UUK cuts of April 2022}
\author[1]{Jackie Grant \thanks{j.j.grant@sussex.ac.uk} }
\author[1,2]{Mark Hindmarsh \thanks{m.b.hindmarsh@sussex.ac.uk}} 
\author[3,4]{Sergey E. Koposov \thanks{skoposov@ed.ac.uk}}

\affil[1]{\textit{Department of Physics and Astronomy, University of Sussex, Brighton BN1 9QH, U.K.}}
\affil[2]{\textit{Department  of  Physics  and  Helsinki  Institute  of  Physics,  PL  64,  FI-00014  University  of  Helsinki,  Finland}}
\affil[3]{\textit{
Institute for Astronomy, University of Edinburgh, Royal Observatory, Blackford Hill, Edinburgh EH9 3HJ, UK}}
\affil[4]{\textit{Institute of Astronomy, University of Cambridge, Madingley Road, Cambridge CB3 0HA, UK}}

\date{May 2022}

\begin{document}

\maketitle

\begin{abstract}
We present the first global analysis of the impact of the April 2022 cuts to the future pensions of members of the Universities Superannuation Scheme. For the 196,000 active members, if Consumer Price Inflation (CPI) remains at its historic average of 2.5\%, the distribution of the range of cuts peaks between 30\%-35\%. This peak increases to 40\%-45\% cuts if CPI averages 3.0\%. The global loss across current USS scheme members, in today's money, is calculated to be {\pounds}16-{\pounds}18 billion, with most of the 71,000 staff under the age of 40 losing between {\pounds}100k-{\pounds}200k each, for CPI averaging 2.5\%-3.0\%. A repeated claim made during the formal consultation by the body representing university management (Universities UK) that those earning under \pounds 40k would receive a ``headline'' cut of 12\% to their future pension is 
shown to be a serious underestimate for realistic CPI projections. 
\end{abstract}

\newpage

\section{Introduction}

The Universities Superannuation Scheme (USS) is a {\pounds}90 billion pension scheme for hundreds of thousands of staff at around 70 ``pre-1992'' universities in the UK, and many other smaller academic and related institutions. 
In the last decade scheme members have seen a succession of significant cuts to their pensions \cite{Grove_2015, PLATANAKIS201614}.  
We survey the effect of the most recent cuts, which came into force at the beginning of April 2022.

Using only data issued by USS and Universities UK (UUK), we calculate total future pension loss due to the original UUK proposal for cuts, announced April 2021. We refer to this as the ``UUK proposal''.
This proposal was used in the on-line modeller \cite{USS_consultation_Nov_2021} produced by USS as part of the formal consultation with scheme members, which we use to calculate projected pensions under the pre- and post-April 2022 rules.\footnote{An adjusted UUK proposal, announced in February 2022, and imposed on 1 April 2022, delayed the CPI cap by two years. We demonstrate in the appendix that 
cuts modelled under the adjusted proposal differ by a maximum of 1.2 percentage points from those under the original for our range of CPI.
}

\begin{figure}[hb!]

\centering

\includegraphics[scale=1.0]{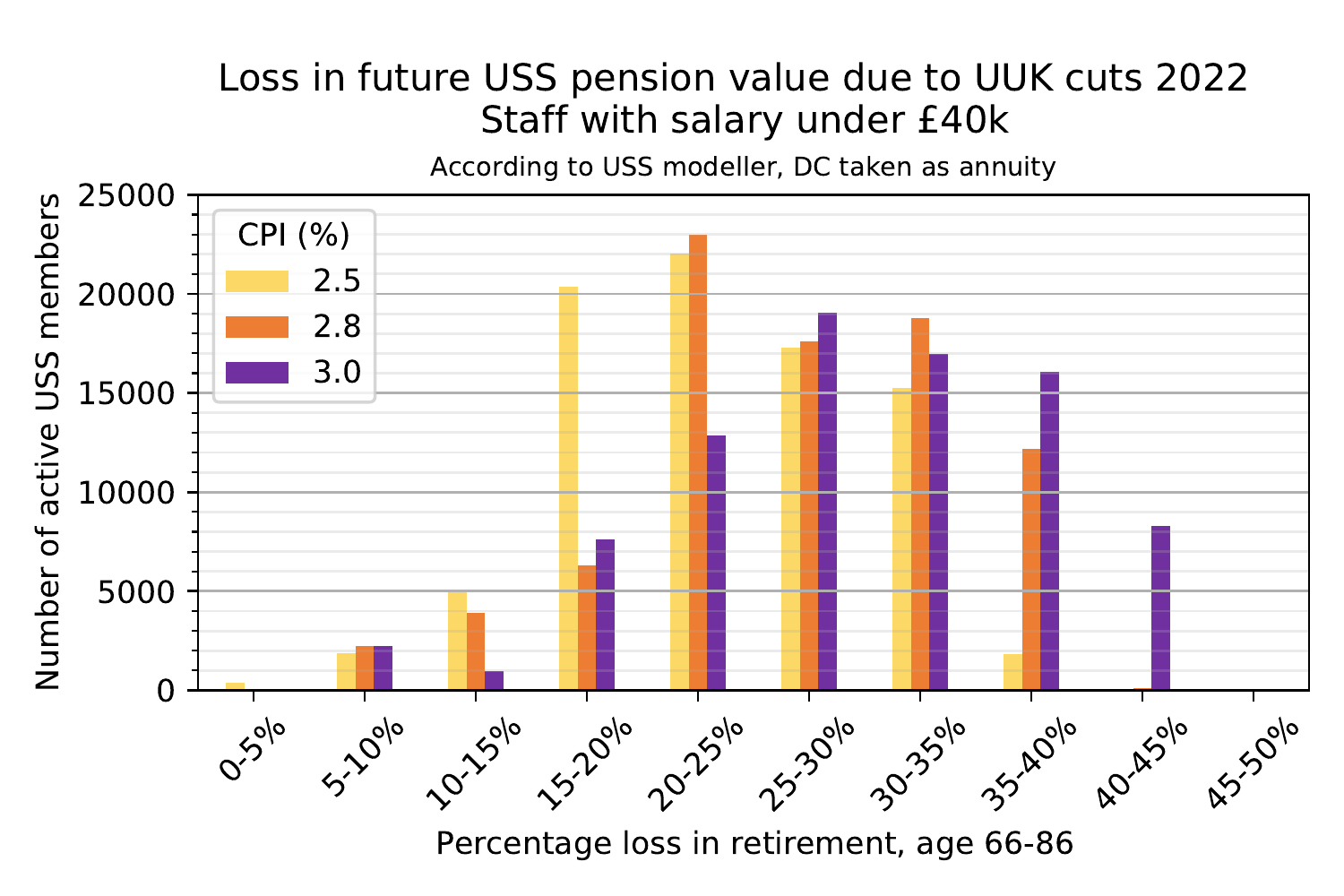}

  \caption{Number of current staff in USS who earn below {\pounds}40,000 (84,000 of 196,000) against percentage loss to future pension under the UUK proposal. Total pension is calculated as DB+DC as annuity. The distribution is shown for CPI projections averaging 2.5\%, 2.8\% and 3.0\%. All analysis uses USS and UUK data only.}
    \label{fig:cpi all 40k}

\end{figure}

The pension scheme, which until 2011 was a final salary Defined Benefit (DB)\footnote{See Table \ref{tab:glossary} of glossary in appendix for definition of key terms.} scheme, is now more complex, and so the USS modeller offers a number of options. We detail our choices in the methodology: here we highlight the choice of projected CPI.
%
%
We take a realistic range,   
including the USS projection of March 2021 of 2.5\% (also the historic average, 1989 to date), USS's updated projection of 2.8\%, 
and a rate of 3.0\%.
With these choices, we calculate the loss in future pension, both as a percentage and as a monetary value, across all staff currently in USS. All monetary values given by the USS modeller are in ``today's'' money, that is, adjusted by the assumed rate of CPI.

The UUK proposal reduced accrual in the Defined Benefit (DB) component from 1/75 to 1/85, and lowered the upper salary limit for the DB component from {\pounds}60,000 to {\pounds}40,000. 
Above this threshold, pension payments go towards a 
Defined Contribution (DC) component. 
The proposal included a ``hard cap'' of 2.5\% on the annual indexation of the DB pension value. 
Under the previous ``soft cap'' indexation, 
the DB pension value was fully matched to CPI in the range 0-5\%, with half matching of the excess above 5\%, up to a maximum increase of 10\%.
Note that even if the mean CPI is below the hard cap, there will be some years in which the cap is exceeded, and 
the pension value will still fall behind inflation. We will see that this is an important effect.


A motivation for our work was 
the UUK claim, published 28 May 2021, that a 
\textit{... staff member earning under the salary threshold of {\pounds}40,000 per annum, the UUK proposal would lead to a headline reduction of about 12\% in future pension benefits}  \cite{THE_12_28_May_2021, UUK_12_May_21}. 

Considering only staff in USS who earn below {\pounds}40,000 \footnote{The Heat map omits technical details but appears to show actual salary rather than FTE so may skew towards those closer to retirement but on higher salaries in part-time roles. The percentage loss could therefore be an underestimate for the group who have below {\pounds}40k salaries on a FTE basis. }, Figure \ref{fig:cpi all 40k} shows that under the UUK proposal 90\% (76,800 out of 84,000) lose more than 15\% of their future pension, with percentage loss peaking between 20\%-30\% and extending to 40\%-45\% for realistic values of CPI. 
Our results therefore refute the UUK claim. 

As a consistency check we compare our results with those of the USS personas of Aria, Bryn and Chloe \cite{USS_personas_7_Dec_2021}, finding excellent agreement (see Table \ref{tab:personas Aria, Bryn, Chloe} of the appendix).
We also use the USS personas to explore the loss in total pensions. We find agreement with earlier work analysing cuts to total pensions from a sample of 42 staff \cite{Mirams_Jan_2022} which cast further doubts on UUK claims \cite{OtsukaM_10-18_Dec2021} 
about total pension cuts. 

In November 2021 the university sector embarked on industrial action over USS that continues to date. Despite widespread opposition, the adjusted UUK proposal was passed, having gained employers' support through UUK's formal consultations, by the casting vote of the Chair at the Joint Negotiating Committee on 22 February 2022. 

This paper is organised as follows: this short summary and brief description of the method are followed by key results, comparison of our results with public claims, then conclusions and open questions. A set of appendices detail method, the effect of the CPI caps, a sensitivity analysis and links to all data and code.

\section{Methodology}

The analysis uses two sources of widely available UUK and USS data. The UUK Heat map of active membership numbers by age and salary \cite{USS_Heatmap_7_April_2021} is combined with data on projected pension values taken directly from the USS consultation modeller \cite{USS_consultation_Nov_2021}. 

The modeller allows anyone, whether registered as a guest or a USS member, to compare their projected pension value at ages 66 and 86, with and without the proposed cuts. This pension value is further split as pension earned to April 2022 and future pension, that is, pension projected to be earned between April 2022 and retirement, which we took to be at age 66. As global data on staff pensions already earned is not available, data collected was by necessity for future pensions only; however, this is the most relevant measure for understanding the long-term impact of the proposed changes. The key inputs to the modeller are shown in Table \ref{tab:modeller inputs}
\begin{table}[h]
    \centering
\begin{tabular}{|l|r|r|}

\hline
\textbf{Input}& \textbf{Default}  & \textbf{Used}   \\
\hline
Date of birth  &   &  1 October Y \\
Salary  & & S\\
CPI & 2.5\% & 2.5\%-3.0\% \\
Future salary growth per year & 4.0\% & 4.0\% \\
DC component options on retirement & Drawdown & Annuity \\
Stock market growth per year for DC component & 4.77\% & 4.77\%\\
\hline

\end{tabular}

\caption{Key inputs to USS consultation modeller. We varied Y (year of birth) and S (salary) across the full range of the UUK Heat map of current staff in USS. We chose CPI values of 2.5\%, 2.8\% and 3.0\%.}
    \label{tab:modeller inputs}
\end{table}

Building on earlier work investigating the USS modeller \cite{KoposovS_USS_modelling}, we chose the default values for future salary growth and DC growth rate. We varied the date of birth and salary across the mid-points of the full range of Heat map values. 
The mean value of future CPI can be set by the user (between 0\%-5\%) and we chose 2.5\%, 2.8\% and 3.0\%. For the DC component there was the option to ``drawdown'', ``convert to an annuity'' or take as ``cash''. The annuity option was chosen for all data to best compare, like-for-like, the guarantee of any DB pre-cuts to post-cuts. 
It is then straightforward to determine the loss in retirement income due to the UUK proposal, as the annual retirement income given at ages 66 and 86 can be used to calculate the total loss over 20 years of retirement. 

Application of the Heat map to this loss in future pensions given by the USS consultation modeller then produces a distribution of cuts to future pensions across staff currently in the USS pension scheme. We consider loss to future pensions across all 196,000 staff in USS. We separately consider the effects on the 84,000 staff earning below {\pounds}40k, and the 71,000 staff below the age of 40. Further details on data, method and consistency checks are included in the appendix.

\section{Results}

We present the results as data tables and stacked histograms. The data tables show percentage and monetary loss, as given by the USS modeller, for each of the entries on the USS Heat map. A section of this data is shown in the appendix with links to full tables over the Heat map values. This data is then used to consider the distribution of cuts first grouping by salary and then grouping by age. 

\begin{figure}[H]
\centering

\includegraphics [scale=0.9]{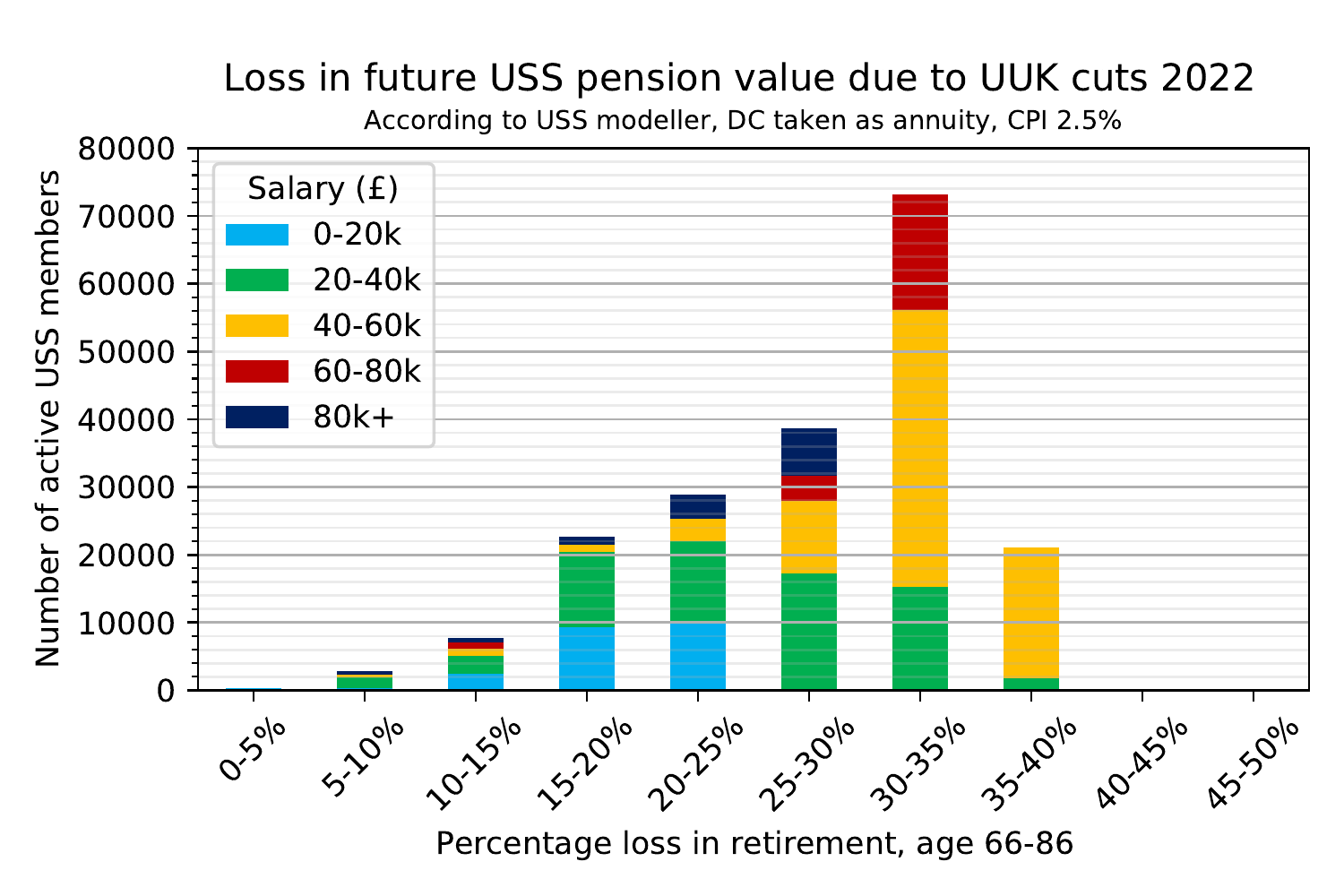}

  \caption{For CPI 2.5\%, percentage loss of future pensions, grouped by salaries, due to USS modelling of UUK cuts when considering DB+DC as annuity. The only three data points USS present are the ``personas'' Aria (37 yrs, {\pounds}30k), Bryn (43 yrs, {\pounds}50k), and Chloe  (51 yrs, {\pounds}70k), these are in the respective ranges 25-30\%, 35-40\%, 30-35\%.
}
    \label{fig:cpi 2.5 salary stacked}

\end{figure}

\begin{figure}[H]
\centering

\includegraphics [scale=0.9]{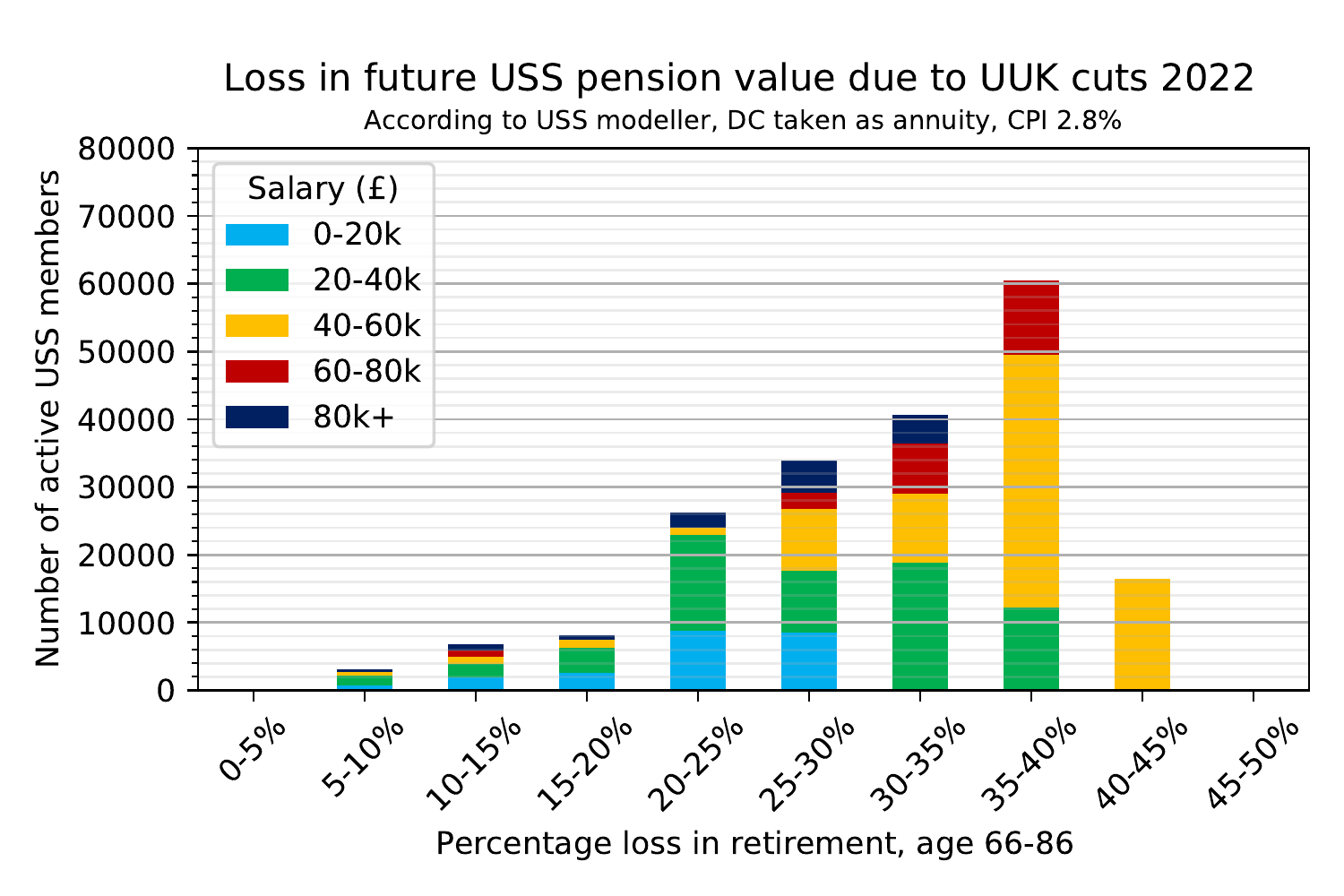}
\includegraphics [scale=0.9]{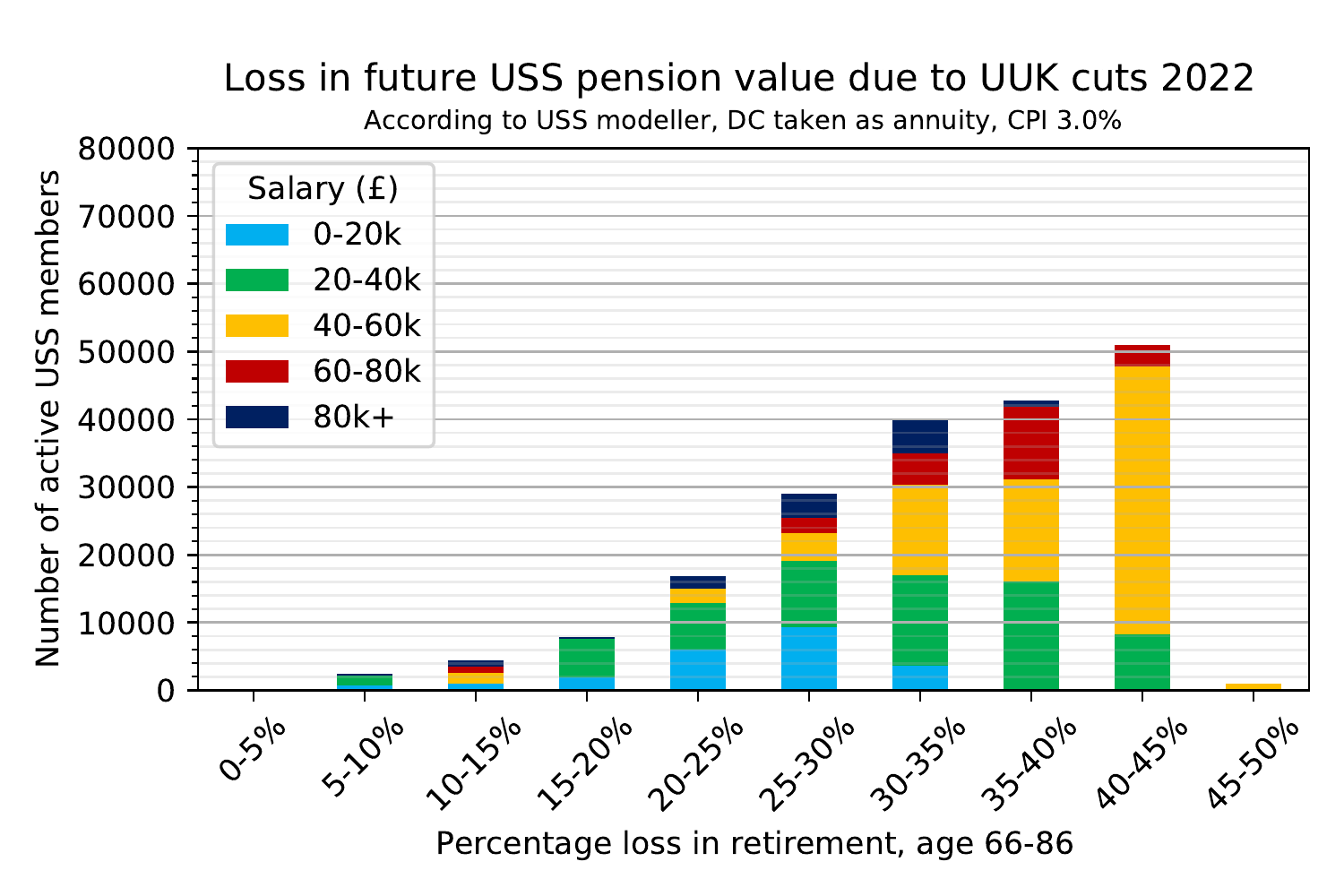}

  \caption{For CPI 2.8\% (top) and 3.0\% (bottom), percentage loss of future pensions, grouped by salaries, due to USS modelling of UUK cuts when considering DB+DC as annuity.  Aria (37 yrs, {\pounds}30k), Bryn (43 yrs, {\pounds}50k), and Chloe  (51 yrs, {\pounds}70k) are in the respective ranges 25-30\%, 35-40\%, 30-35\% (top) and 30-35\%, 35-40\%, 35-40\% (bottom).}
    \label{fig:cpi 2.8 3.0 salary stacked}

\end{figure}

For grouping by salary, the results are shown as histograms for CPI at 2.5\%, 2.8\% and 3.0\% in Figures \ref{fig:cpi 2.5 salary stacked} and \ref{fig:cpi 2.8 3.0 salary stacked}. Along with Figure \ref{fig:cpi all 40k} these figures clearly refute the UUK claim discussed in the summary, that those on salaries below {\pounds}40,000 suffer ``headline'' cuts of 12\%, however one interprets the word headline. The mean values of cuts across all staff are 27\%, 31\% and 33\% with the distribution peaking at 30-35\%, 35-40\% and 40-45\% respectively for CPI of 2.5\%, 2.8\% and 3.0\%.

\begin{figure}[H]
\centering

\includegraphics [scale=0.9]{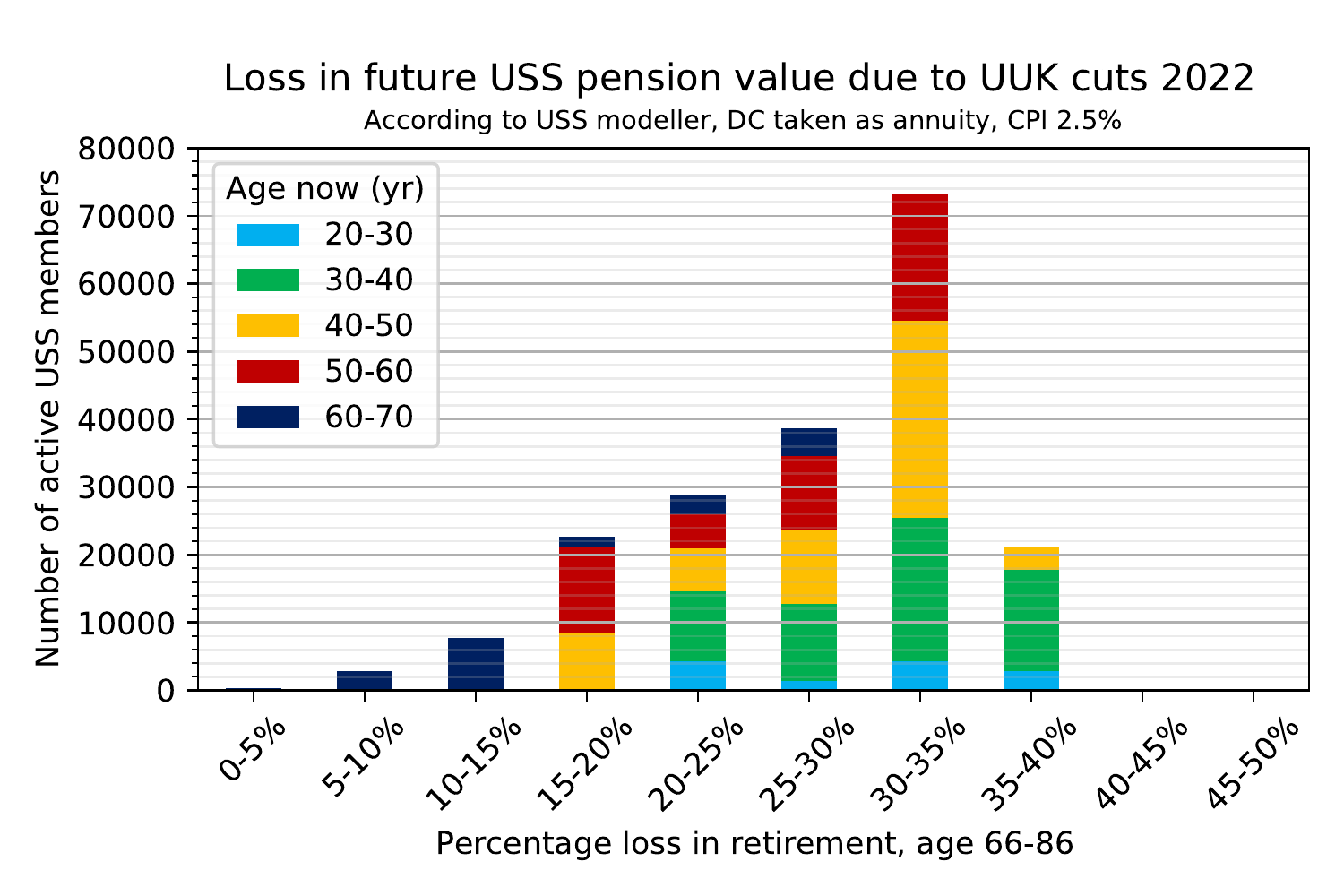}

  \caption{For CPI 2.5\%, percentage loss of future pensions, grouped by age, due to USS modelling of UUK cuts when considering DB+DC as annuity. }
    \label{fig:cpi 2.5 age stacked}

\newpage

\end{figure}

\begin{figure}[H]
\centering
\includegraphics [scale=0.9]{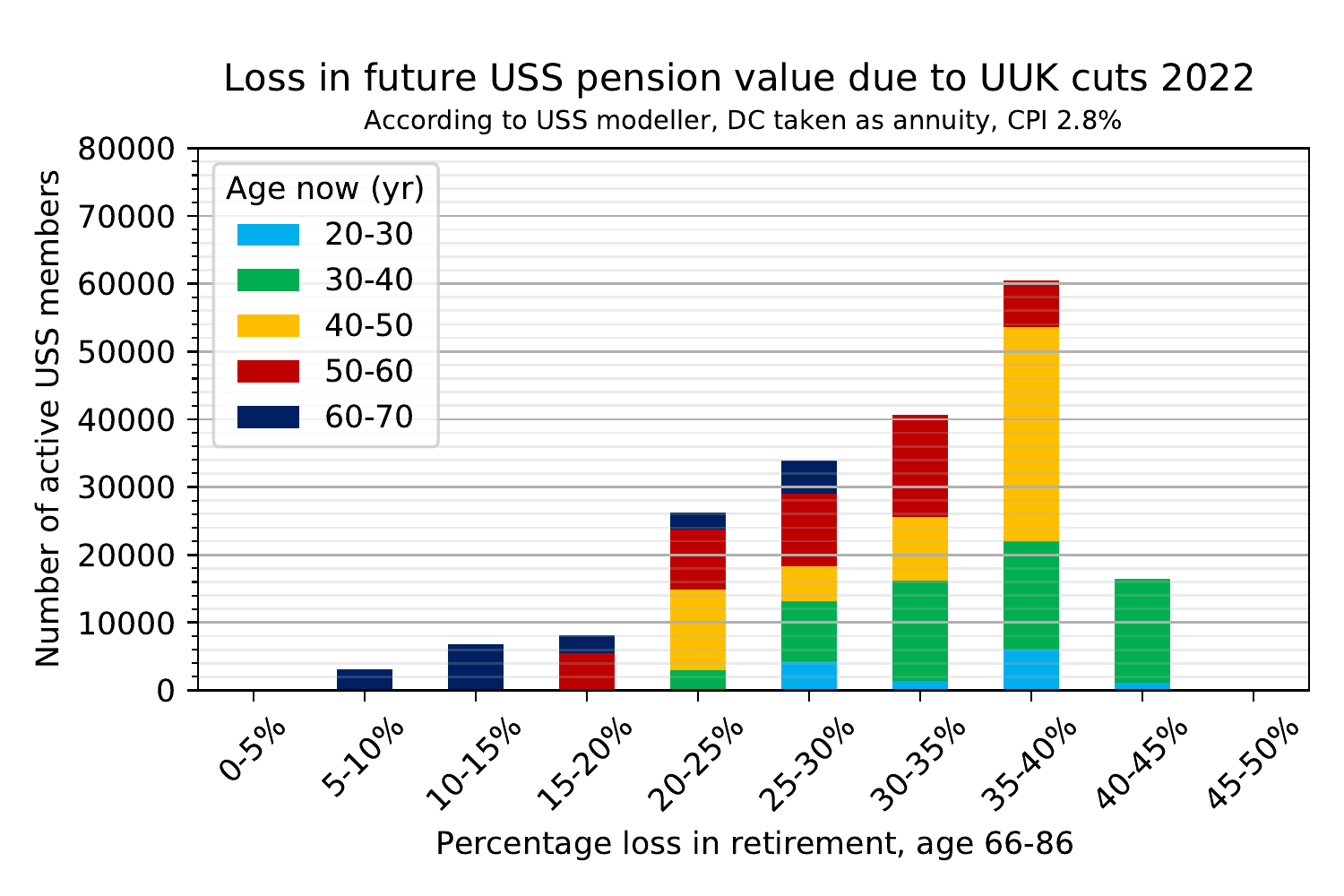}
\includegraphics [scale=0.9]{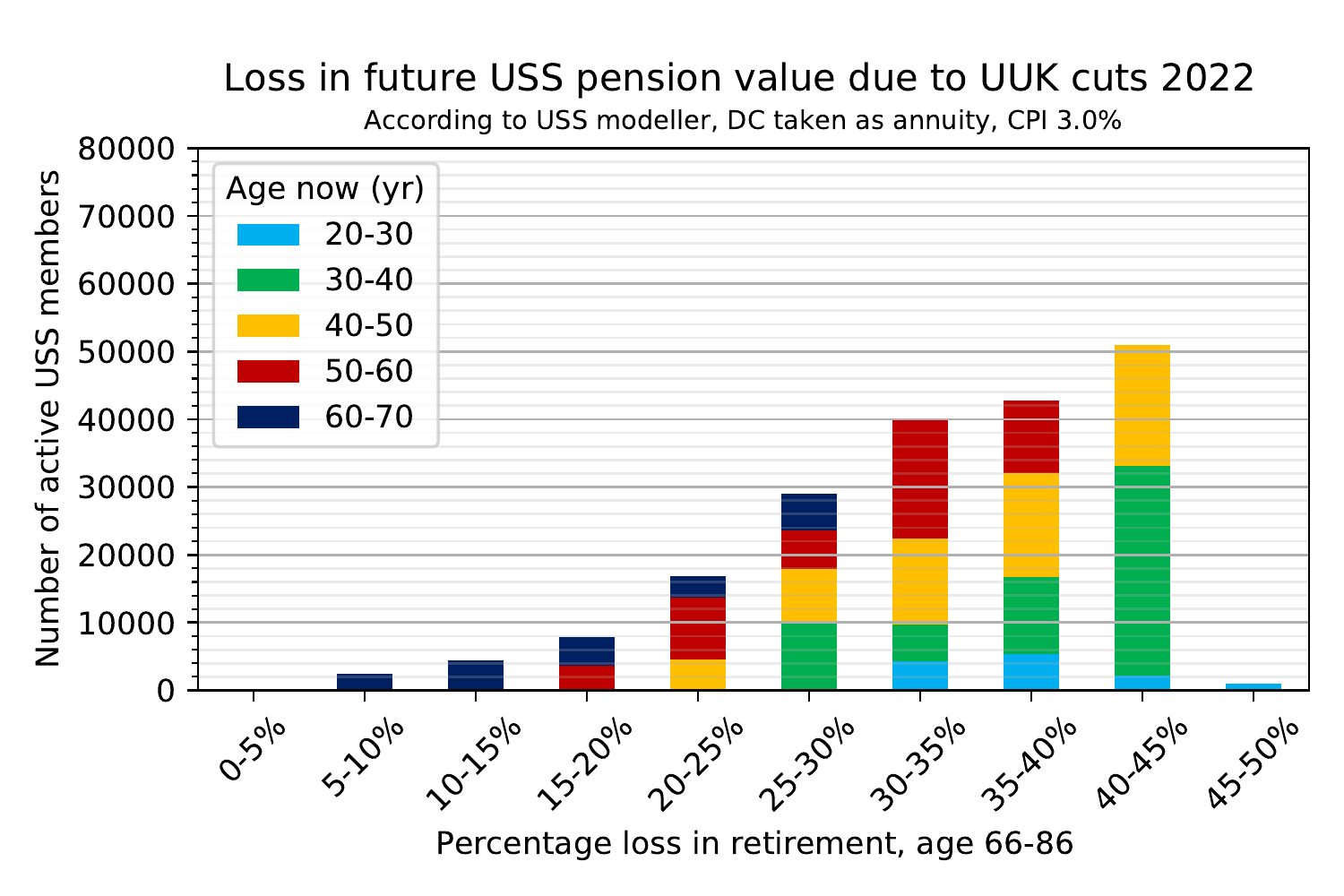}

  \caption{For CPI 2.8\% (top) and 3.0\% (bottom) percentage loss of future pensions, grouped by age, due to USS modelling of UUK cuts when considering DB+DC as annuity.}
    \label{fig:cpi 2.8 3.0 age stacked}
    
    \end{figure}

For grouping by age the results are similarly displayed for CPI at 2.5\%, 2.8\% and 3.0\% in Figures \ref{fig:cpi 2.5 age stacked} and \ref{fig:cpi 2.8 3.0 age stacked}. Table \ref{tab:median} shows how the cuts are consistently larger for those under the age of 40. These differences are significantly larger than our uncertainties, as discussed in the appendix. For CPI 2.8\%, it is staff under the age of 40 who receive the highest percentage of cuts, 40\%-45\%, with the range of cuts for the same age-group peaking between 35\%-40\%. If CPI averages 3.0\%, some staff under 40 experience cuts in the range 45\%-50\%. 

\begin{table}[H]
    \centering
\begin{tabular}{|l|r||r|r|r||r||r|}

\hline
\textbf{Group} & \textbf{CPI} & \textbf{Q1 }& \textbf{Q2} & \textbf{Q3} & \textbf{Mean}  & \textbf{Mode range}      \\
\hline
\hline
All & 2.5\% & 22\% & 29\% & 33\% & 27\% & 30\%-35\%  \\
staff & 2.8\% & 26\% & 33\% & 37\% & 31\% & 35\%-40\%  \\
 & 3.0\% & 28\% & 35\% & 40\% & 33\%  & 40\%-45\% \\
\hline
\hline
Staff & 2.5\% & 26\% & 32\% & 35\% & 30\% & 30\%-35\%  \\
$<$ 40 & 2.8\% & 30\% & 37\% & 40\% & 35\% & 35\%-40\%   \\
yrs old & 3.0\% & 33\% & 39\% & 43\% & 38\%  & 40\%-45\% \\
\hline
\end{tabular}

\caption{Interquartile ranges for percentage loss to future pension, with DC as annuity. Q1, Q2, Q3 represent quartile ranges for 25\%, 50\% and 75\% of the membership. Mean and mode range are also shown. The top half of the table includes all 196k current staff, the lower half of the table is all 71k current staff below the age of 40. }
    \label{tab:median}
\end{table}

\newpage

The loss in monetary value can similarly be calculated and Table \ref{tab:monetary loss global} shows the global loss,  calculated across all current staff in the USS scheme, as  {\pounds}16-{\pounds}18bn in today's money, for our range of projected CPI. If we consider only those 71,000 staff below the age of 40 (making up 36\% of the staff in USS) we find that just over 50\% of the cuts fall on their pensions. 
Table \ref{tab:median money} shows the mean loss across the whole membership as {\pounds}82-{\pounds}94k per person, with the interquartile ranges Q1, Q2 and Q3 showing a large variation. The mean loss rises to {\pounds}121-{\pounds}139k per person when considering only those staff below the age of 40. When considering all staff, we see that most staff receive cuts of less than {\pounds}50k, but for staff under the age of 40, most staff receive cuts between {\pounds}100-200k.

\begin{figure}[H]

\includegraphics [scale=0.9]{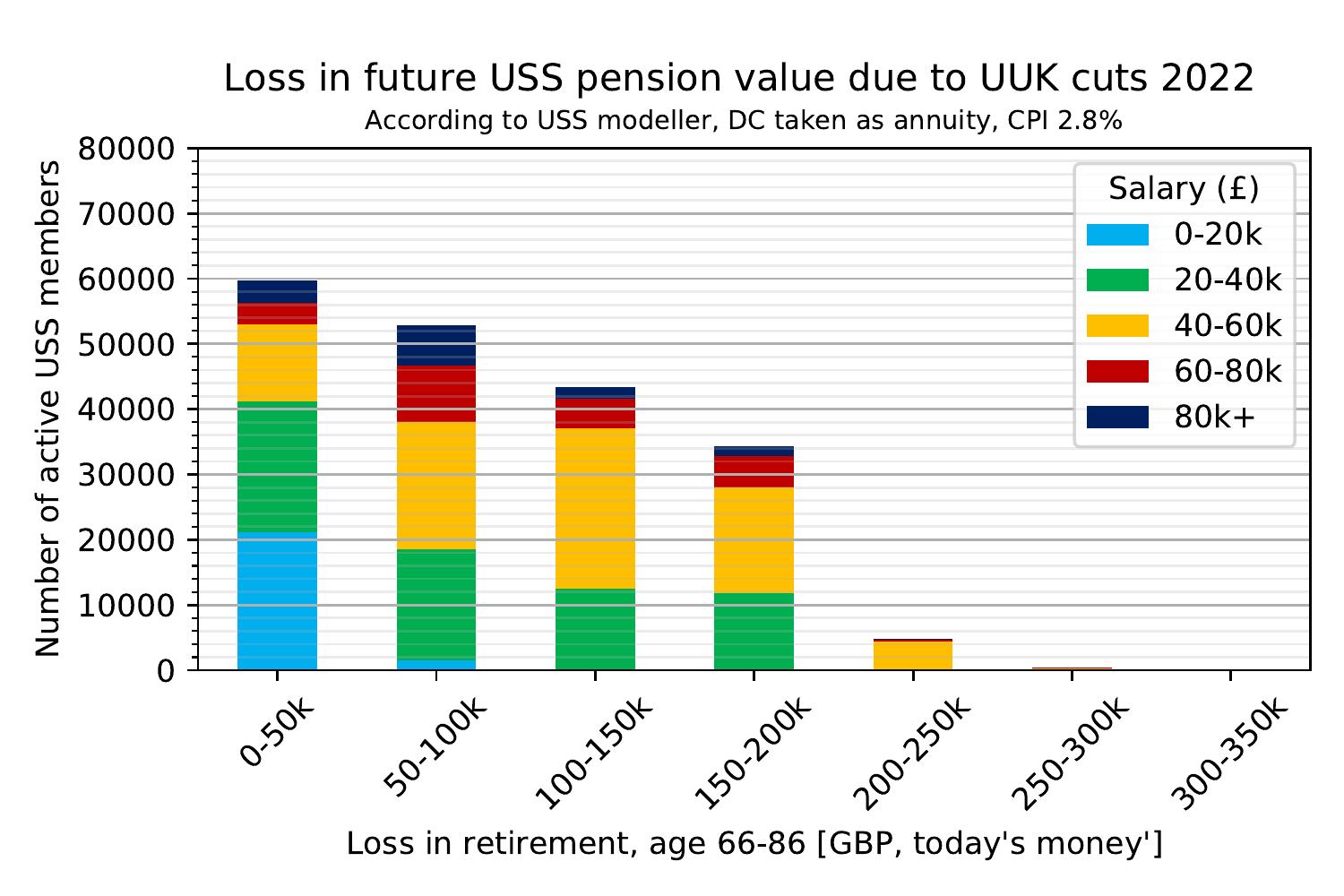}
\includegraphics
[scale=0.9]{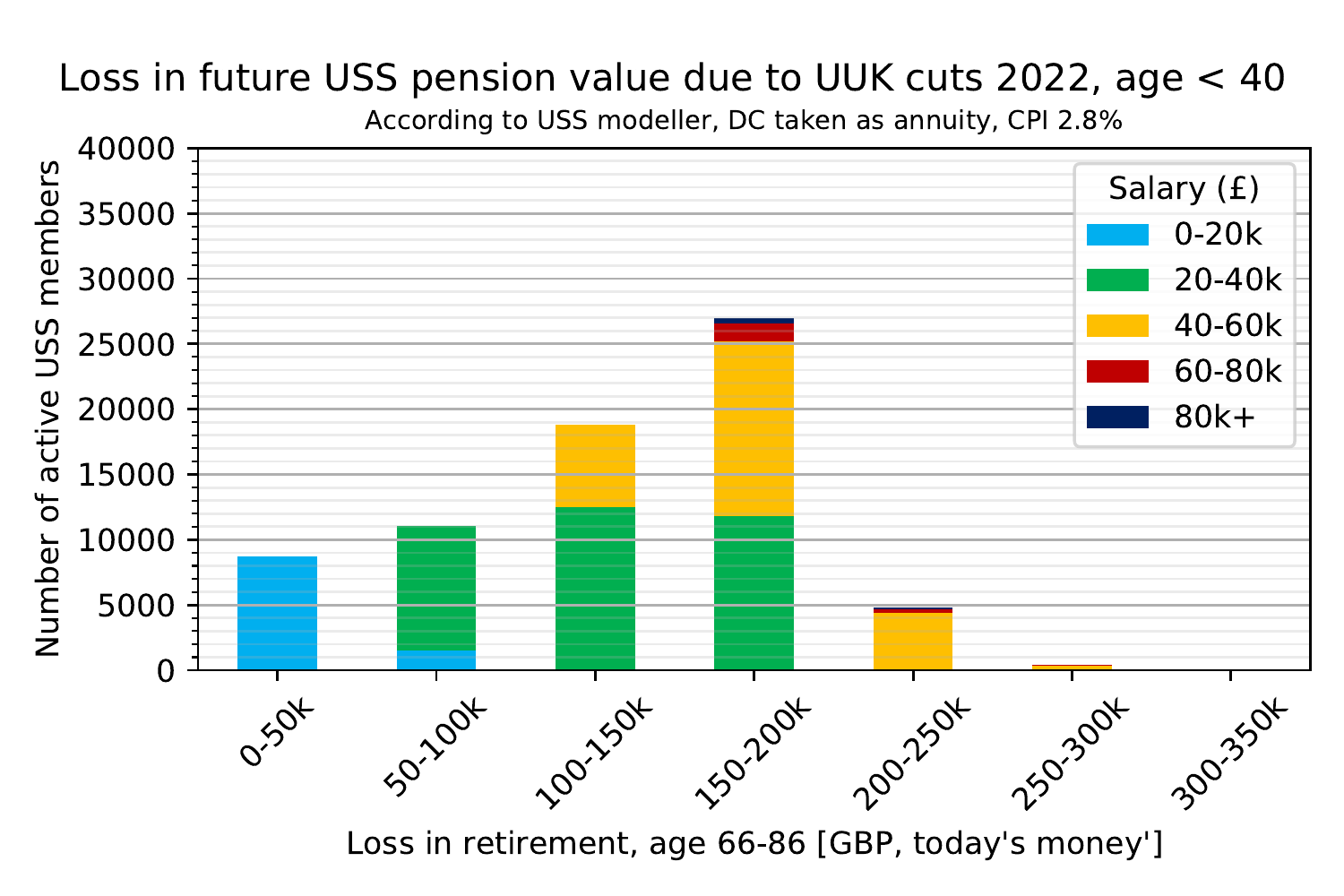}
  \caption{For CPI 2.8\%, monetary loss in today's money, grouped by salary, due to USS modelling of UUK cuts when considering DB+DC as annuity. The top graph shows monetary loss, in today's money, across the whole membership, the lower graph considers only those USS members below the age of 40. }
    \label{fig:cpi 2.8 money loss age stacked}

\end{figure}

Figure \ref{fig:cpi 2.8 money loss age stacked} (top) 
shows the distribution of cuts across all staff, binned by monetary value. It demonstrates that 
over 80,000 people receive cuts of between {\pounds}100-{\pounds}300k in today's money, for USS's current CPI projection of 2.8\%.  Figure \ref{fig:cpi 2.8 money loss age stacked} (bottom) shows the distribution of cuts binned by monetary value for those under 40, giving a clear representation of the larger cuts in monetary value experienced by younger staff.

\begin{table}[h]
    \centering
\begin{tabular}{|c||c||c|}

\hline
\textbf{CPI} & \textbf{Global loss }& \textbf{Global loss }    \\
\textbf{projection}  & \textbf{all 196k staff}& \textbf{All 71k staff $<$ 40 yrs} \\
\hline
\hline
2.5\% & {\pounds}16.1 bn & {\pounds}8.5 bn  \\
2.8\% & {\pounds}17.6 bn & {\pounds}9.4 bn  \\
3.0\% & {\pounds}18.4 bn & {\pounds}9.8 bn  \\
\hline

\end{tabular}

\caption{Monetary loss across current scheme membership of 196,000 staff, and the 71k current staff below the age of 40, CPI-adjusted to today. Global loss is calculated by combining the UUK Heat map of USS scheme membership by salary and age with the results of the USS consultation modeller.}
    \label{tab:monetary loss global}
\end{table}

\begin{table}[H]
    \centering
\begin{tabular}{|l|r||r|r|r||r|r|}

\hline
\textbf{Group}& \textbf{CPI} & \textbf{Q1 }& \textbf{Q2} & \textbf{Q3} & \textbf{Mean} & \textbf{Mode range}    \\
\hline
\hline
All & 2.5\% & {\pounds}31k & {\pounds}78k & {\pounds}125k & {\pounds}82k & {\pounds}0k-50k \\
staff & 2.8\% & {\pounds}36k & {\pounds}85k & {\pounds}149k & {\pounds}90k & {\pounds}0k-50k  \\
 & 3.0\% & {\pounds}38k & {\pounds}89k & {\pounds}140k & {\pounds}94k & {\pounds}0k-50k  \\
\hline
\hline
Staff & 2.5\% & {\pounds}81k & {\pounds}136k & {\pounds}157k & {\pounds}121k & {\pounds}100k-150k  \\
$<$ 40 & 2.8\% & {\pounds}91k & {\pounds}149k & {\pounds}173k & {\pounds}133k  & {\pounds}150k-200k  \\
yrs old& 3.0\% & {\pounds}97k & {\pounds}154k & {\pounds}181k & {\pounds}139k & {\pounds}150k-200k \\
\hline
\end{tabular}

\caption{Interquartile ranges for monetary loss to future pension, with DC as annuity. Q1, Q2, Q3 represent quartile ranges for 25\%, 50\% and 75\% of the membership. Mean and mode range are also shown. All figures CPI-adjusted or in today's money. The top half of the table includes all current staff, the lower half of the table is all 71k current staff below the age of 40.}
    \label{tab:median money}
\end{table}

\section{Comparison of UCU and UUK claims to our results}

Conflicting claims have been made regarding the level of cuts. Our work is, to our knowledge, the first global analysis of cuts to future pensions, and so we can compare the claims to our results.

On 28 May 2021, UCU launched their own modeller to illustrate the level of cuts, the accompanying press release claimed 
\begin{quote}
\textit{A typical member on a lecturer's salary, aged 37, will suffer a 35\% loss to the guaranteed retirement benefits which they will build up over the rest of their career.} UCU, 28 May 2021 \cite{UCU_May_2021}
\end{quote}

This was later clarified to mean a salary of {\pounds}39-41k. 
Our data, extracted from the USS modeller, shows that a member of staff aged 37.5 earning {\pounds}42.5k, will lose 34-38\% of their future pension for CPI at 2.5-2.8\% (see Table \ref{tab:percentage loss heatmap}, and associated data files). Our results are therefore compatible with this statement. 

On the same day as the UCU press release, UUK responded with their own press release, critical of the UCU modeller, which included the 
claim about the headline reduction of 12\% 
for those earning under {\pounds}40k, 
which we discussed in the introduction. 

This UUK claim, supporting UUK's proposal, was challenged
at the time \cite{OtsukaM_28_May_21} by demonstrating that a 12\% reduction in future pension was entirely accounted for by the accrual rate change and effectively ignored the impact of the 2.5\% CPI cap. We discuss in the appendix why such a CPI cap imposes the most detrimental loss of the UUK proposal's three changes. 

UUK maintained this claim, without supporting evidence or qualifying comment, throughout formal consultations with employers. The UUK claim of a future reduction of 12\% for staff earning below {\pounds}40k remains unevidenced and unqualified, in at least two places \cite{UUK_12_May_21,UUK_back_changes_15_June_2021}, on the UUK website dedicated to these USS consultations. 

In December 2021, UUK commented on the cuts to total pensions, stating that

\begin{quote}
\textit{Modelling published by the USS Trustee themselves} [Aria, Bryn, Chloe] \textit{shows that impact} [of UUK's proposed pension cuts] \textit{is likely to be between 10\% and 18\% for the vast majority of members.} UUK, 1 December 2021 \cite{OtsukaM_Fitt_Nov2021}
\end{quote}

These figures refer to the range across the three USS personas. 
The details of these three personas can be manually entered into the USS modeller and the cuts to their future pensions compared with the corresponding data of the Heat map values that we extracted from the USS modeller (see 
Table \ref{tab:personas Aria, Bryn, Chloe}). 
We consider the cuts in total and future pension received by the three USS personas, with the DC taken as both drawndown and annuity across the CPI range 2.5-3.0\%. Using our data for salaries and ages across the Heat map, we obtain excellent agreement with the manual entering of data. The manual entering of data across the CPI range supports the view that UUK were underestimating the cuts to total pensions for the three personas. 
The underestimate arises in considering only the cuts at age 66 (so omitting the further erosion after the age of 66) and using a low estimate for the annual devaluation due to CPI fluctuations. 

The UCU General Secretary wrote to UUK in November 2021 asking for clarification over misleading statements on the level of cuts, and the UUK Chief Executive replied rejecting the claim that statements were misleading \cite{GradyJ_Nov2021, UUK_Jarvis_Nov_2021}.

\section{Conclusions and open questions}

We have performed the first global analysis of cuts to future pensions of USS members as a consequence
of the UUK cuts of April 2022. 
We use only data from the USS consultation modeller, and the UUK Heat map of current staff numbers by salary and age. 
The pension consists of a Defined Benefit (DB) component and a Defined Contribution (DC) component. To best compare the cuts to the pension, we assumed that the DC component is converted into an annuity. 
We consider the results across a realistic range of CPI values.

Our results lead to a three main conclusions. 

First, the cuts are significant and, because of the introduction of a hard cap of 2.5\% on pension indexation, 
sensitive to CPI projections and assumptions. The cuts to future pensions
can be quantified in various ways. 
For example, the interquartile range for percentage loss to future pension is 22-40\% across the whole membership.  The mean loss, in today's money, is £82-94k per person, with an interquartile range of £31-140k across CPI values of 2.5-3.0\%. 
The loss across all current 196k USS members is £16-18bn, for the same CPI range.
The hard cap of 2.5\%  accounts for the largest part of the cuts to future DB pensions, 22-35\% during 20 years of retirement, following a career of 40 years for CPI averaging between 2.5\% and 2.8\%.

Second, if we consider only the 71k staff below the age of 40, the interquartile range for percentage loss is higher, 26-43\%. Because the percentage cuts are larger for younger staff and because they have more years of work ahead, this represent a much larger loss in today's money, with mean values of loss between £121-139k per person, and an interquartile range £81-181k for CPI varying between 2.5-3.0\%. 

Finally, we compared our results to public claims on the level of cuts from UUK and UCU. Our results are compatible with UCU's claim of 35\% loss to future guaranteed pension of staff on a typical lecturer's salary. However, we found the UUK claim of headline 12\% cut to future pension of those earning below £40k to be a significant underestimate. Our consistency checks also supports the view that UUK's claim on level of cuts to total pensions of between 10-18\% is 
based on three data points, the USS personas Aria, Bryn and Chloe, and is also an underestimate. 

The differences between the UUK claims and our calculations raises certain questions. 

UUK is a body representing Vice-Chancellors of universities, and the official organisation that consults employers and their governing bodies on changes to the scheme. It is not known whether UUK carried out this kind of global analysis; however, UUK does have the capability. The data we have used has been available to USS and UUK for over a year, and UUK has a large team of researchers with specialism in pensions and finance. If they did not carry out this analysis, should they have commissioned such work? If they did carry out this analysis, then why was it not shared publicly?

In addition to the public statements, university governing bodies were provided with confidential briefings from UUK and we have no way of knowing the analysis contained in these briefings. Given that the cuts were imposed following a series of consultations by UUK, during which  underestimates of the cuts were repeated, we ask: were governing bodies were given sufficiently accurate information to assess the impact of the cuts to current and future staff?  

We thank Antonio Padilla and Michael Otsuka for helpful comments on drafts of this manuscript. Computations and plots were performed using pandas \cite{jeff_reback_2022_6408044}, NumPy \cite{2020NumPy-Array},  and Matplotlib \cite{Hunter:2007}.

\bibliographystyle{unsrt}
\bibliography{USS_bib.bib}

\newpage

\appendix
\section{Appendix}

This appendix covers the method for collecting and analysing all data, a discussion of the UUK proposals, investigation of impact of the CPI caps and sensitivity analysis. 
All data are available at \href{https://github.com/SussexUCU/USS/tree/main/data/USS_modeller}{https://github.com/SussexUCU/USS/tree/main/data/USS\_modeller} along with calculations and code. We include a short glossary and present tables of the UUK Heat map, and future loss in percentage and pounds across all values for CPI at 2.8\%.

\subsection{Method}

\subsubsection*{What is the USS modeller?}
The USS modeller \cite{USS_Modeller_2021}, launched on 5 November 2021\footnote{It was relaunched eight days later, on correction of the annual devaluation from 0.35\% to 0.50\% \cite{OtsukaM_10-18_Dec2021}},
provided information to USS members as part of the legally required consultation on the UUK proposal. The stated aim was to \textit{`help you see how your benefits and contributions to USS could change as a result of the proposals'.} The USS modeller's assumptions were agreed by UUK.

Shortly after the launch of the USS modeller, UUK produced guidance \cite{UUK_modeller_guidance_12_Nov_2021} for USS members on using the modeller. They encouraged use of the modeller by stating that the \textit{`consultation is about members and their representative bodies understanding the impacts of the proposed changes'}. In this spirit, and building from earlier work \cite{KoposovS_USS_modelling},  we collected data, as described in Table \ref{tab:Modeller data}, from the USS modeller across the range of salaries and ages of the UUK Heat map \cite{USS_Heatmap_7_April_2021} of USS active membership. 

\subsubsection*{How we got the data} 
To extract the results of the USS forecasts we did the following. 
Using the python module Selenium\footnote{\href{https://www.selenium.dev/}{https://www.selenium.dev/}} and the web driver geckodriver\footnote{https://github.com/mozilla/geckodriver}, we wrote a python script to interact with the USS modeller \href{https://www.ussconsultation2021.co.uk/}{https://www.ussconsultation2021.co.uk/}. 
We used a local copy of the USS modeller, as all
the calculations are performed in the browser, without any interaction with the server.
The code automates the interaction with the browser, such as clicking ``pop-up'' and ``submit'' buttons,
sets input values for the modeller's JavaScript, and collects the modeller's outputs.
The inputs include the date of birth of the person, salary, inflation,
salary increase, and investment growth values.
The outputs we collected were the annual pension under the 2021 scheme and the UUK proposal scheme at retirement and 20 years after retirement. Table \ref{tab:Modeller data} contains a summary of inputs and outputs.
Note that we did not collect the lump sum.  As explained in the next section,  this causes a small underestimation in the calculation of the size of cuts. 

We executed the code for an array of dates of birth and salaries, according to the UUK Heat map grid give in Table \ref{tab:UUK heatmap}. The code of the python scripts, together with the USS pages are available on request.

\begin{table}[h]
    \centering
\begin{tabular}{|l|l|}

\hline
\textbf{Data input} &  \textbf{Details} \\
\hline
 Date of birth &  1 October in year given by mid-point of Heat map values. \\
 Salary &  Salary on 1 April 2022, mid-point of Heat map values.\\
 Salary growth & Nominal salary growth over career, default 4.0\%. \\
 CPI & Average CPI over a career, 2.5\% (default), 2.8\%, 3.0\%. \\
 DC component &``Annuity'' chosen over ``cash'' and ``drawdown''.  \\
 DC growth & Default nominal 4.77\% return on the stock market. \\
 \hline
\textbf{Data collected} &  \textbf{Details} \\
\hline
2021 pension pa, age 66 &  Annual pension received aged 66 under 2021 scheme. \\
 UUK pension pa, age 66 & Annual pension received aged 66 under UUK proposal.\\
 2021 pension pa, age 86 & Annual pension received aged 86 under 2021 scheme. \\
 UUK pension pa, age 86  & Annual pension received aged 66 under UUK proposal.\\
 \hline
\end{tabular}
\caption{Table of data-types input and collected from USS modeller. The annual pension received aged 66 and 86 is CPI-adjusted, or in USS's terminology, valued in ``today's money''. The DC component was chosen as an annual annuity to best compare like-for-like between the scheme, as it was to April 2022, and the UUK proposal. There was a choice of average CPI and the modeller included an annual CPI devaluation to present the impact of variance under the UUK proposed hard cap of 2.5\% CPI. This annual devaluation is discussed in the section on CPI caps.}
    \label{tab:Modeller data}
\end{table}

\begin{figure}[h]

\includegraphics [scale=0.5]{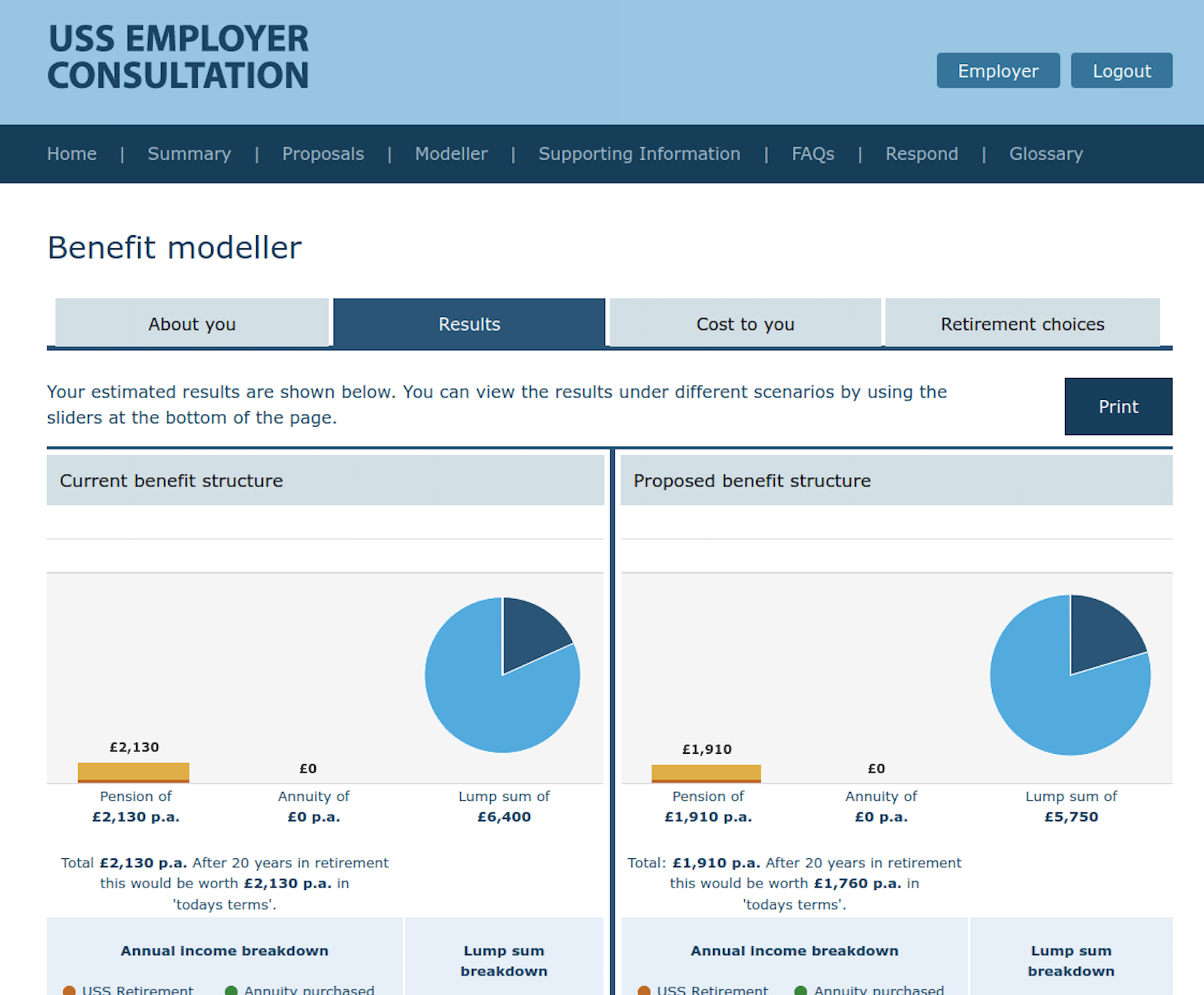}

  \caption{The USS consultation modeller in display mode. ``Current benefit structure'' is pre-April 2022, ``Proposed benefit structure'' represents the projected pension value  under the UUK proposal.}
    \label{fig:modeller}

\end{figure}

\subsubsection*{How we calculated loss to future pensions} 
To perform the calculation of retirement income between ages 66 and 86, from the USS modeller data given at ages 66 and 86, we proceeded as follows. 

The DC component is assumed to be converted into an annuity (a fixed annual payment) with a USS-specified conversion rate, whose numerical value is in the range 35-45 years$^{-1}$ for retirement aged 66.  The modeller divides the total DC savings component by this annuity conversion to get an annual fixed income from the DC component.
The annuity is assumed by USS to be indexed with CPI: therefore a linear interpolation between the income at the two ages gives the income at all years in between. 

Under the UUK proposal, the DB component devalues every time CPI fluctuates above 2.5\%.  The average effect of this fluctuation is accounted for by an annual devaluation assumption, and so pension income is modelled with a geometric progression, whose start and end points are given by the modeller pension values at ages 66 and 86.

Across the membership values of age and salary, a geometric progression will slightly over-estimate pension income for those on high salaries, due to the proposed {\pounds}40,000 DB/DC threshold, while a linear interpolation will slightly under-estimate pension income for those on low salaries. We chose a linear interpolation for all income, but ran consistency checks with a geometric progression to verify that the linear interpolation used was slightly underestimating the cuts. 
Both the future percentage loss and the future monetary loss were calculated. 

In choosing salary and date of birth input values, to best represent the ranges of the Heat map, we chose the mid-point values with two exceptions. 

The highest salary of the Heat map is {\pounds}150k+ with no upper bound. To estimate a mid-range value for this salary value we consulted the Annual Statements from 2019 and 2020 of the University of Sussex, as a representative University. We considered those staff paid over {\pounds}150k+, and calculated the mean as {\pounds}200,000. So we used {\pounds}200,000 as the representative salary for those classified by the Heat map as {\pounds}150k+. 

Similarly for the age value 65+, again given with no upper bound, we took the age 65.5, as this is six months from the retirement date we chose. We also ran a check for age 67.5, which would have given zero loss, and found a negligible difference in results.

\subsection{ACAS 2018, the UUK proposal, and the UUK adjusted proposal}

The USS valuation of 31 March 2020 was viewed by many experts, including UUK's own actuary, as excessively pessimistic. 
The pessimism had two components: the undervaluing of the scheme's assets just after the start of the Covid pandemic, and a belief that markets would grow at well below historic long-term rates. \cite{Rau_R_historic_returns,UUK_letter_USS_March2021, UUK_prop_7_April_2021, Wolf_FT_2021, Woon_W_suprlus}.

The UUK proposal to severely cut USS pension benefits was announced on 7 April 2021, within a UUK employer consultation on USS that ran to 24 May 2021 \cite{UUK_prop_7_April_2021}. This UUK proposal is an almost identical (but more expensive) version of the ACAS agreement between UUK and UCU of March 2018 \cite{ACAS_March_2018}. The ACAS agreement was rapidly rejected by UCU and replaced with an offer that maintained the status quo, while a new valuation was completed and a Joint Expert Panel convened \cite{JEP_March_2018}.

\begin{table}[h]
    \centering
\begin{tabular}{|c||c||c|c|c|}

\hline
& \textbf{USS}& \textbf{ACAS} &  \textbf{UUK } & \textbf{UUK adjusted}  \\
& \textbf{2021} & \textbf{proposal 2018} & \textbf{proposal 2021} & \textbf{proposal 2022}  \\
\hline
\hline
Accrual rate & 1/75 & 1/85 & 1/85 & 1/85  \\
\hline
DB/DC  & $\pounds$60k & $\pounds$42k & $\pounds$40k & $\pounds$40k  \\
\hline
CPI cap & soft cap  & hard cap & hard cap  & hard cap  \\
Full indexation & 0-5\% & 0-2.5\% & 0-2.5\%  & 0-2.5\%  \\
+ 1/2 indexation & 5-15\% & none   & none & none \\
Hard cap delay & n/a & none   & none & two years \\

\hline

Employer & 21.1\% & 19.3\% & 21.1\% & 21.3\% \\
Employee &9.1\% & 8.7\%   & 9.1\% & 9.1\% \\
\hline

\end{tabular}

\caption{USS scheme structure pre-April 2022, alongside the 2018 ACAS proposal, the 2021 UUK proposal and the 2022 UUK adjusted proposal. The UUK adjusted proposal was imposed on 1 April 2022. Employer and employee contributions are included.}
    \label{tab:UUK and UUK adjusted proposals}
\end{table}

Returning to the UUK proposal of April 2021, following the reporting on the legally required consultation \cite{OtsukaM_consultation_responses,USS_consultation_responses}, UUK put forward an adjusted proposal \cite{UUK_adj_prop_02_Feb_2022} for consultation with employers in February 2022. 
The accompanying technical note on the adjusted proposal \cite{USS_tech_note_UUKadj_02_Feb_2022} demonstrates that the adjustment amounts to delay of the CPI hard cap by two years, as the CPI cap comes into effect in April 2024, and is then implemented in April 2025 for the entire DB component earned since April 2022. 

We quantify the small difference in loss between the UUK proposal and the adjusted proposal in the next section, where we discuss the important effect of the CPI cap.

\subsection{Discussion of impact of CPI caps}\label{subsec:cpi}

Even when the mean CPI is below the hard cap of 2.5\%, CPI fluctuates around its mean and the pension is  devalued, in real terms, each time the value of CPI goes above 2.5\%. In the USS modeller, this effect is taken into account by an annual devaluation assumption for each projected mean CPI.

One way to compare the impact of the hard cap and soft cap on the indexation of a pension scheme is to consider the impact of these various projections and assumptions used by UUK and USS since the valuation date of March 2020. 

As given by Eq.~(\ref{e:AnnDev}) one of the UUK assumptions was a value for the CPI adjustment, $a$, of 0.5\%. This adjustment uniquely sets the annual devaluation, $d$, for each value of projected CPI, $c$, given a CPI hard cap $h$. 

\begin{align}
1-d &=\frac{(1+h-a)}{(1+c)} 
\label{e:AnnDev}
\end{align}

This annual devaluation models the impact of the variance of CPI above the hard cap of 2.5\% for the USS modeller. The value of $h$ is 2.5\% for the UUK proposals and $c$ varies between 2.5\%-3.0\% for our analysis. UUK assumed $a$ of 0.5\% and this is the value used in the USS modeller. USS has stated they would have used $d$ of 0.58\% for projected CPI of 2.5\%, which, using the same approach, allows us to calculated the USS assumption for the CPI adjustment, $a$, as 0.59\%.

\begin{table}[h]
    \centering
\begin{tabular}{|l||r|r|r|r|r|}

\hline
 & \textbf{2020} & \multicolumn{2}{|c|}{\textbf{2021}} & \multicolumn{2}{|c|}{\textbf{2022}}  \\
\hline
\hline
USS CPI projection (c) &  2.1\% &  \multicolumn{2}{|c|}{2.5\%} & \multicolumn{2}{|c|}{2.8\% }\\
\hline
\hline
UUK devaluation used for USS modeller (d)  &   & 0.5\% &   & 0.8\% & \\
USS devaluation assumption (d) &  0.4\% & & 0.58\% & & \textit{0.87\%} \\
\hline
\hline
20 years of 2.5\% hard cap erosion  & -8\% & -10\% & -11\% & -15\%  &  \textit{-15\%}\\
40 years of 2.5\% hard cap erosion  & -15\% & -18\% & -21\% & -27\%   & \textit{-29\%}\\
60 years of 2.5\% hard cap erosion  & -21\% & -26\% & -29\% & -38\%   & \textit{-41\%} \\
\hline
Average loss from CPI hard cap ages 66-86 & -18\% & -22\% & -25\% & -33\% & \textit{-35\%} \\
\hline

\end{tabular}

\caption{Pension value erosion due to the 2.5\% hard cap on indexing of Defined Benefit pensions introduced by the UUK proposal.
The columns show various UUK and USS assumptions for CPI projections and the associated annual devaluation. The rows show the cumulative effect on the value of the DB pension over various time periods. The far-right column data, given in italics, is calculated using the same method as the UUK devaluation calculation that reproduces UUK assumption values. 
}
    \label{tab:annual devaluation UUK USS}
\end{table}

Taking the average retirement as 20 years and estimating working life to be 40 years, Table \ref{tab:annual devaluation UUK USS} shows that, simply by compounding the devaluation over the given number of years, the expected erosion from the UUK hard cap of 2.5\% ranges between 22-35\% for UUK and USS's own assumptions since the launch of the modeller. 

The accrual rate drop from 1/75 to 1/85, using the standard percentage calculation, corresponds to a 12\% reduction. The drop of the threshold from {\pounds}60k to {\pounds}40k is difficult to quantify, as it depends on the DC fund growth rate and annuity price assumptions, but it will only impact the window of salary between {\pounds}40-60k. 

It is clear that, of the three changes of the UUK proposals (threshold drop, accrual drop and CPI hard cap) the 2.5\% hard cap has the largest impact. Further, that the impact of the CPI cap is highly sensitive to projected CPI assumptions. 

The USS projection for CPI, originally 2.5\% when the modeller was launched in November 2021, was updated to 2.8\% in January 2022. The UUK assumption for annual devaluation of 0.5\% for CPI at 2.5\%  is smaller than the USS assumption of 0.58\%. We consider the impact of this higher projection for CPI and the higher USS assumption for annual devaluation further in the sensitivity analysis. 

In the rest of this section, we consider the difference between the UUK proposal and the UUK adjusted proposal which, as explained in the previous section, is a delay in the imposition of the CPI hard cap by two years from April 2022 to April 2024.

One can quantify the impact of a two year delay on the percentage loss from one year's contribution to future pensions. This has been calculated, using USS modelling assumptions, for a 40-year-old on {\pounds}40k and amounts to less than one percentage point reduction in loss \cite{OtsukaM_UUKadjusted_2_Feb_2022}. As discussed in the sensitivity analysis, applying the same technique globally across the UUK Heat map produces similarly small change in the percentage loss, 0.6-1.2 percentage points, see Table \ref{tab:annual devaluation}, which is within the sensitivity of our results.

It could be argued that the delay of the hard cap is currently more valuable than average, as we are now experiencing high values of CPI.
However, it will not change the conclusion that, of the three parts of the UUK proposal, the 
introduction of a CPI hard cap results in the most significant loss to 
scheme members in the long term. This is shown in Table \ref{tab:annual devaluation UUK USS}. As discussed in the next section, it will also not change the conclusion that the difference between the results of the UUK proposal and UUK adjusted proposal are insignificant when using the annual devaluation approximation method of the USS modeller that aims to represent the long-term losses.

\subsection{Sensitivity analysis and consistency checks}

As demonstrated in the discussion above, the CPI cap has the largest impact of the three parts of the UUK cuts to the scheme imposed on 1 April 2022. 
The cuts are therefore highly sensitive to the assumed CPI, as can be seen in the shift of the peak of percentage cuts by 10 percentage points for changes in CPI from 2.5\% to 3.0\%, shown in Figures \ref{fig:cpi 2.5 age stacked} and \ref{fig:cpi 2.8 3.0 age stacked}. It can also be seen in the increase in global loss from {\pounds}16.1 to {\pounds}18.4 bn, shown in Table \ref{tab:monetary loss global}, for CPI increasing from 2.5\% to 3.0\%. With the imposition of the CPI hard cap of 2.5\% the defined benefits have become highly uncertain in real terms. 

We considered four further areas: (i) precision of the USS modeller (ii) use of linear or geometric interpolation (iii) DC growth and annuity assumptions (iv) annual devaluation assumptions due to CPI fluctuations.  

Our analysis suggests that quoting results to one percentage point is within the precision of our calculations. The highest sensitivity is on CPI projections with DC growth and annuity conversion also showing high sensitivity.

\textbf{(i) The precision of the USS modeller.} 

The value of pensions at the age of 66 and 86 are rounded to the nearest {\pounds}10, meaning the pension value could be $\pm${\pounds}5. The lowest value of pension output by the modeller is {\pounds}1,100 meaning the uncertainty is less than 0.5\%. For a 40 year old on {\pounds}40k, the associated uncertainty is then less than 0.05\%. 
Rounding therefore does not introduce significant uncertainty into the results. 

\textbf{(ii) Use of a linear interpolation between ages of 66 and 86.}

The annual devaluation of the DB component of the pension compounds losses in a geometric series. The DC component is assumed to be constant in real terms. 
However, we did not record the DB and DC components separately, and so a geometric interpolation between income at 66 and 86 would slightly under-estimate the total pension income, while linear interpolation would slightly over-estimate it. 
The difference is larger for lower salaries, where the DB component dominates. For example, 
for salaries below {\pounds}40k, the difference between the interpolations is around 0.05-0.1\% disappearing to 0.01\% for the highest salaries. This is well within the precision of the results we quote. 

\textbf{(iii) DC growth and annuity conversion assumptions.}

These will only impact on salaries above the DB/DC threshold of {\pounds}40k. 
If USS DC growth assumptions are varied so that on retirement the annuity is $\pm$10\% of the 
default, 
the difference in pension loss for a 40-year-old on {\pounds}60k is $\mp$6 percentage points. We calculate this from the loss due to one years contributions, see \href{https://github.com/SussexUCU/USS/tree/main/data/USS_modeller}{github}. The dependence on the DC assumptions of growth and future annuity cost are therefore potentially significant and deserves further consideration beyond the scope of this paper.

\begin{table}[h]
    \centering
\begin{tabular}{|l | c||c|c|c|}

\hline
\textbf{CPI} & \textbf{Hard cap} & \textbf{Loss from UUK} &   \textbf{Loss from USS}  & \textbf{Difference}  \\
 &  \textbf{delay} & \textbf{devaluation}  &     \textbf{devaluation} & \textbf{percentage point} \\
\hline
\hline
\multirow{2}{*}{2.5\%} & none &  27.2\% & 28.7\%  &  1.5\% \\
 & 2 years  & 26.6\% & 28.0\% &   1.4\%\\
\hline

\hline
& difference  & 0.6\% & 0.7\% &   \\
\hline
\hline
\multirow{2}{*}{2.8\%}  &  none & 32.0\% & 33.4\% & 1.4\%  \\
& 2 years & 31.1\% & 32.4\% &1.3\% \\ 
\hline

\hline
& difference   & 0.8\% & 1.0\% &   \\
\hline
\hline
\multirow{2}{*}{2.8\%} & none & 35.0\%  & 36.3\%  & 1.3\% \\
 & 2 years & 33.9\%  & 35.1\% & 1.2\% \\
\hline

\hline
& difference  & 1.1\% & 1.2\% &   \\
\hline
\hline

\end{tabular}

\caption{Global percentage loss in retirement income from one year's contribution due to UUK cuts using both the UUK devaluation assumption of 0.5\% and the the USS annual devaluation assumption of 0.59\%. The difference in percentage loss for these two assumptions varies between 1.2 and 1.5 percentage points. The delay of the CPI cap by two years is also included, where the difference in loss with and without the two year delay is between 0.6 and 1.2 percentage points.}
    \label{tab:annual devaluation}
\end{table}

\textbf{(iv) Assumption for annual devaluation due to CPI fluctuations.} 

The USS modeller compares benefits under the scheme in 2021 to benefits under the UUK proposal, with UUK  assumptions. The UUK assumption for the devaluation impact of CPI was used for the USS modeller. This UUK assumption produced a smaller devaluation that the USS assumption. We attempt to quantify the difference between the UUK and the USS assumption for devaluation as follows. 

It is possible, using the assumptions of the USS modeller, to calculate the global percentage loss under both the UUK devaluation assumption and the more pessimistic USS devaluation assumption. Table  \ref{tab:annual devaluation} shows the results for CPI projected between 2.5\% to 3.0\%. To calculate the difference we calculated the percentage loss due to one years contribution across the whole active membership, weighted by the UUK Heat map. We did this both with the UUK assumption and the USS assumption.

Before considering consistency checks including the lump sum sensitivity, we briefly consider the pay growth assumption. We have not performed analysis on the pay growth assumption of 4\%, although our work could readily be extended in this way. We only note for now that a nominal pay growth of 4\% on average over a career is not an unreasonable assumption. Also that a smaller pay growth would be expected to reduce the pre-April 2022 projected pension and a larger pay growth to increase the pre-April 2022 projected pension. Therefore pay growth increases could be expected to increase the loss to projected pension and pay growth decreases to decrease the loss.

Finally we turn to consistency checks. The raw data from the USS modeller was cross-checked with data from USS personas Aria, Bryn and Chloe giving excellent agreement. 

We entered the data for the USS personas Aria, Bryn and Chloe manually recording all the outputs for CPI at 2.5\%, 2.8\% and 3.0\%. This allows a calculation of the percentage loss to future pension both with and without the lump sum, and showed that omitting the lump sum results in an underestimate of the future pension loss by at most 1 percentage point for our range of projected CPI. 

This work also allowed for consistency checks between manually recording data from the USS personas and our method for recording and calculating the impact of the UUK proposals. The UUK Heat map values do not exactly coincide with the salaries of the three personas and so we interpolated between the two nearest points from the Heat map data, either side of the personas. the consistency checks results are shown in Table  \ref{tab:personas Aria, Bryn, Chloe} along with the UUK claims of loss to total pension. 

Because USS states pensions already earned by the three personas it is straightforward to include a check on the loss to total pension as well as future pension. Losses to total pension for the three personas has been considered earlier \cite{Mirams_Jan_2022, OtsukaM_Fitt_Nov2021} and we expand by including considering a loss to total pension, both with drawdown and annuity options for the DC fund across our range of values for CPI. The UUK claim on loss to total pension of between 10-18\% is, again, shown to be an underestimate. We note that UUK figures of 10-18\% are calculated using the outdated devaluation assumption and by only considering the cuts at the age of 66, thereby omitting any erosion during retirement. 

 The choice of ages and salaries of the three personas Aria, Bryn and Chloe are also noteworthy. The UUK Heat map, Table  \ref{tab:UUK heatmap}, shows that for people aged 35-39 (which includes Aria's aged of 37) the most common salary is £40-45k, however Aria was chosen with a much lower salary of £30k. Had Aria been chosen to have the most common salary for her age the cuts to their future pension would be 38\% rather than 27-31\% for CPI at 2.8\%. However the most common salary for someone aged 40-44 (which includes Bryn’s age of 43) is lower than Bryn’s stated salary of £70k. The same is true for the age range 50-54, which includes Chloe’s age of 51. By choosing the youngest persona to have a salary lower than the most common salary, while both the older personas have salaries higher than the most common salaries, the representative cuts for younger staff are underplayed and level of intergenerational inequality of the global data is obscured.

All the analysis and plots were produced independently by authors from the raw data and gave identical results. Raw data, code, checks and calculations are all available at \href{https://github.com/SussexUCU/USS/tree/main/data/USS_modeller}{https://github.com/SussexUCU/USS/tree/main/data/USS\_modeller} along with calculations and code

\begin{table}[H]
    \centering
\begin{tabular}{|c|r||r|r|r|r|r||r||r|}

\hline
\small \textbf{USS} & \small \textbf{CPI} & \small \textbf{UUK} &  \multicolumn{2}{|c|}{\small\textbf{Modeller+drwdwn}} & \multicolumn{2}{|c|}{\small\textbf{Modeller+annty}} & \small\textbf{Av.} & \small\textbf{Our}\\

\small\textbf{prsna}   & & \small\textbf{claim} & \small\textbf{tot. loss}  & \small\textbf{fut. loss}  & \small\textbf{tot. loss} & \small\textbf{fut. loss}  & \small\textbf{loss}  & \small\textbf{value} \\
\hline
\hline
\cellcolor{red!25}Aria & 2.5\% & 16\% &19\%-26\% & 21\%-28\% & 20\%-27\%  &  22\%-29\% & 26\%  & 25\%\\
\cellcolor{red!25}37yrs & 2.8\% & & 22\%-31\% &24\%-34\% & 23\%-32\% & 25\%-36\% & 30\% & 29\% \\
 
\cellcolor{red!25}{\pounds}30k & 3.0\% & & 23\%-35\% &26\%-38\% & 24\%-36\% & 27\%-39\% & 33\% & 33\% \\
 \hline
 \hline
\cellcolor{green!25}Bryn & 2.5\% & 18\% & 23\%-27\% &29\%-34\% & 27\%-31\% & 34\%-39\%  & 37\% &  35\% \\
\cellcolor{green!25}43yrs & 2.8\%  & & 25\%-31\%  &32\%-39\% & 29\%-35\% & 37\%-45\% & 41\%& 39\%\\  
\cellcolor{green!25}{\pounds}50k & 3.0\%  & & 26\%-33\% & 34\%-42\% & 31\%-37\% & 39\%-48\% & 43\%& 41\%\\
 \hline
  \hline
\cellcolor{blue!25}Chloe & 2.5\% & 10\% &  13\%-15\%& 25\%-30\% & 15\%-17\% &31\%-35\% & 33\% & 31\%\\
\cellcolor{blue!25}51yrs & 2.8\% & &  14\%-17\% & 28\%-34\% &16\%-19\% & 33\%-40\%& 36\% & 34\% \\
\cellcolor{blue!25}{\pounds}70k & 3.0\% & &  14\%-18\% & 29\%-37\% &17\%-21\% & 35\%-43\%& 39\% & 36\% \\
 \hline

\end{tabular}
\caption{Percentage loss for three USS personas Aria, Bryn and Chloe taken by entering their salaries and ages by hand into the USS modeller. Results shown for total pension loss (tot. loss) and future pension loss (fut. loss) with the range given between ages 66 and 86. The DC options are given as both drawdown (modeller+drwdwn) and annuity (modeller+annty). The two columns on the right show the average loss to future pension with DC as annuity (av. loss) and our calculations (our value).  To calculate our value we took two data points either side of each of the three personas. The results show excellent agreement with a slight underestimate as predicted in our method. The figures in the UUK claim column are the loss to total pension quoted by UUK, the figures are calculated for loss at age 66 only for CPI of 2.5\% using out-dated, lower annual devaluation assumptions \cite{OtsukaM_Fitt_Nov2021}.}
    \label{tab:personas Aria, Bryn, Chloe}
\end{table}

\subsection{Glossary}

\begin{table}[H]
    \centering
\begin{tabular}{|l|p{0.9\textwidth}|}

\hline
\textbf{Terms used} & \textbf{Description} \\
\hline
\hline
CPI  &  Consumer Price Inflation, or CPI is one of the Office for National Statistics (ONS) price indices \cite{ONS_CPI}. It measures the price increase of ``shopping baskets'' of goods and services, so is one measure of the cost of living. CPI has been recorded since 1989 and data is available monthly, quarterly and yearly. Discussion and consultation on price indices is ongoing, and the Royal Statistical Society has a statement on RPI and other indices \cite{RSS_2018}.  Criticisms of the standard basket include undervaluing the index for low-income groups \cite{Guardian_Vimes_Boots_26_Jan_2021}.  \\
\hline
Accrual rate  & The accrual rate is usually given as a fraction such as 1/75 or 1/85. It is the rate at which  pension scheme members earn their pension. For example if the accrual rate is 1/80 and a person pays into the scheme for 40 years, they will earn 40/80 of the final pension, or 1/2 the final pension. The final pension may be a Career Average Revalued Earnings (CARE) or Final Salary.  USS is CARE, so for the 1/75 accrual rate you need to work 37.5 years to get half your career average salary in retirement.  
\\
\hline
DB  & Defined Benefit. This means the pension you receive is ``defined'' or guaranteed to be a particular value during your retirement. If it is index linked to CPI it will retain its value in real terms. \\
\hline
DC  & Defined Contribution. Here the pension you receive is not ``defined'' or guaranteed. Instead the amount contributed is defined and that contribution goes to a savings pot that will increase or decrease in value in an undefined way. USS is hybrid DB/DC. Prior to April 2022 salaries below {\pounds}60k were entirely DB, salaries above {\pounds}60k were DB up to {\pounds}60k and DC above. USS has a range of options for their DC component including a default ``life-style'' a ``life-style ethical'' and some ``do-it-yourself' options. On retirement it is then the scheme members' responsibility to decide how to use the DC pot saved up. Options include drawdown (ie taking regular amounts), taking the cash, or converting the DC to an annuity. \\
\hline
Threshold  & In the context of USS this refers to the DB/DC threshold. Above this threshold contributions go to a DC savings pot, and below the threshold to a DB pension scheme. Before April 2022 the threshold was {\pounds}60k and index linked, following the UUK cuts the threshold was lowered to {\pounds}40k and indexed at CPI of 2.5\%.  \\
\hline
Annuity  & On retirement you can convert any DC savings to an annuity. This buys a guaranteed income in retirement. The USS modeller makes assumptions about the future cost of annuities to be able to model how a DC component of a pension could be converted into a regular guaranteed index linked income.  \\
\hline
\end{tabular}

\caption{Glossary of key terms. See also USS glossary \cite{USS_glossary}, USS consultation glossary \cite{USS_cons_glossary} and UUK glossary \cite{UUK_glossary}.  }
    \label{tab:glossary}
\end{table}

\newpage 

\subsection{Tables of UUK Heat map, percentage and GBP loss for CPI 2.8\%}

\begin{table}[!ht]
    \centering
    \begin{tabular}{|r||r|r|r|r|r|r|r|r|r|r|}
    \hline
      
        Age / age &  $\leq$ 25 & 25-29 & 30-34 & 35-39 & 40-44 & 45-49 & 50-54 & 55-59 & 60-64 & 65+ \\ \hline
        \hline
        0–5k & 321 & 1,292 & 1,029 & 711 & 490 & 456 & 428 & 406 & 310 & 300 \\ \hline
        {\pounds}5–10k & 120 & 721 & 893 & 779 & 630 & 526 & 482 & 495 & 451 & 378 \\ \hline
        {\pounds}10–15k & 122 & 483 & 698 & 681 & 697 & 622 & 555 & 562 & 461 & 376 \\ \hline
        {\pounds}15–20k & 154 & 466 & 787 & 971 & 907 & 728 & 673 & 634 & 487 & 367 \\ \hline
        {\pounds}20–25k & 219 & 647 & 979 & 1,276 & 1,213 & 992 & 906 & 801 & 590 & 434 \\ \hline
        {\pounds}25–30k & 381 & 1,209 & 1,388 & \cellcolor{red!25}1,441 & 1,295 & 1,066 & 989 & 843 & 579 & 358 \\ \hline
        {\pounds}30–35k & 305 & 3,588 & 5,863 & \cellcolor{red!25}3,773 & 2,614 & 2,088 & 1,744 & 1,452 & 835 & 384 \\ \hline
        {\pounds}35–40k & 91 & 1,722 & 6,187 & 4,804 & 2,908 & 1,950 & 1,471 & 1,079 & 643 & 319 \\ \hline
        {\pounds}40–45k & 20 & 658 & 4,283 & 6,333 & 5,017 & 3,673 & 3,058 & 2,231 & 1,166 & 478 \\ \hline
        {\pounds}45–50k & 5 & 247 & 2,446 & 4,751 & \cellcolor{green!25}4,498 & 3,405 & 2,708 & 2,019 & 1,035 & 367 \\ \hline
        {\pounds}50–55k & 0 & 64 & 782 & 2,655 & \cellcolor{green!25}3,415 & 2,667 & 2,376 & 1,902 & 934 & 308 \\ \hline
        {\pounds}55–60k & 0 & 31 & 415 & 1,686 & 2,853 & 2,696 & 2,328 & 1,916 & 969 & 393 \\ \hline
        {\pounds}60–65k & 0 & 16 & 156 & 735 & 1,733 & 2,111 & 2,099 & 1,715 & 907 & 351 \\ \hline
        {\pounds}65–70k & 0 & 7 & 73 & 326 & 802 & 1,047 & \cellcolor{blue!25}1,038 & 868 & 528 & 201 \\ \hline
        {\pounds}70–75k & 0 & 7 & 52 & 233 & 581 & 850 & \cellcolor{blue!25}956 & 842 & 487 & 217 \\ \hline
        {\pounds}75–80k & 0 & 7 & 23 & 122 & 294 & 510 & 663 & 575 & 373 & 179 \\ \hline
        {\pounds}80–85k & 0 & 0 & 19 & 91 & 266 & 352 & 520 & 513 & 310 & 144 \\ \hline
        {\pounds}85–90k & 0 & 0 & 8 & 47 & 181 & 295 & 476 & 545 & 320 & 178 \\ \hline
        {\pounds}90–95k & 0 & 0 & 17 & 56 & 125 & 265 & 316 & 354 & 262 & 147 \\ \hline
        {\pounds}95–100k & 0 & 0 & 10 & 40 & 141 & 194 & 265 & 271 & 213 & 114 \\ \hline
        {\pounds}100–105k & 0 & 0 & 8 & 23 & 80 & 154 & 218 & 289 & 185 & 117 \\ \hline
        {\pounds}105–110k & 0 & 0 & 0 & 16 & 41 & 121 & 162 & 211 & 153 & 64 \\ \hline
        {\pounds}110–115k & 0 & 0 & 11 & 26 & 36 & 71 & 123 & 180 & 142 & 86 \\ \hline
        {\pounds}115–120k & 0 & 0 & 0 & 15 & 30 & 57 & 106 & 124 & 106 & 57 \\ \hline
        {\pounds}120–125k & 0 & 0 & 5 & 14 & 22 & 66 & 102 & 130 & 93 & 52 \\ \hline
        {\pounds}125–130k & 0 & 0 & 0 & 7 & 35 & 37 & 67 & 117 & 76 & 39 \\ \hline
        {\pounds}130–135k & 0 & 0 & 0 & 0 & 23 & 31 & 67 & 70 & 55 & 34 \\ \hline
        {\pounds}135–140k & 0 & 0 & 0 & 6 & 16 & 30 & 49 & 81 & 37 & 31 \\ \hline
        {\pounds}140–145k & 0 & 0 & 0 & 7 & 22 & 25 & 50 & 72 & 38 & 19 \\ \hline
        {\pounds}145–150k & 0 & 0 & 0 & 0 & 10 & 16 & 30 & 61 & 42 & 22 \\ \hline
        {\pounds}150+ & 0 & 0 & 16 & 33 & 65 & 108 & 217 & 309 & 265 & 124 \\ \hline
    \end{tabular}
    
    \caption{UUK Heat map, transposed from the original at \cite{USS_Heatmap_7_April_2021}.  USS's three data points, represented as Aria (37 yrs, {\pounds}30k), Bryn (43 yrs, {\pounds}50k), and Chloe  (51 yrs, {\pounds}70k) are shaded in respectively pink, green and blue. Data as .csv at \href{https://github.com/SussexUCU/USS/tree/main/data/USS_modeller}{github} 
    }
    \label{tab:UUK heatmap}
\end{table}

\begin{table}[!ht]
    \centering
    \begin{tabular}{|r||r|r|r|r|r|r|r|r|r|r|}
    \hline
        Salary\textbackslash age &$\leq$25&25-29&30-34& 35-39 & 40-44 & 45-49 & 50-54 & 55-59 & 60-64 & 65+ \\ \hline
    \hline
        {\pounds}0–5k &  29\% & -28\% & -27\% & -25\% & -24\% & -23\% & -22\% & -19\% & -14\% & -13\% \\ \hline
        {\pounds}5–10k & -29\% & -27\% & -26\% & -25\% & -24\% & -22\% & -21\% & -18\% & -15\% & -11\% \\ \hline
        {\pounds}10–15k & -29\% & -27\% & -26\% & -25\% & -24\% & -22\% & -20\% & -18\% & -15\% & -8\% \\ \hline
        {\pounds}15–20k & -29\% & -28\% & -26\% & -25\% & -23\% & -22\% & -20\% & -18\% & -16\% & -6\% \\ \hline
        {\pounds}20–25k & -32\% & -29\% & -27\% & -25\% & -23\% & -22\% & -20\% & -18\% & -15\% & -7\% \\ \hline
        {\pounds}25–30k & -36\% & -33\% & -30\% & \cellcolor{red!25}-27\% & -24\% & -22\% & -20\% & -18\% & -15\% & -6\% \\ \hline
        {\pounds}30–35k & -39\% & -37\% & -34\% & \cellcolor{red!25}-31\% & -28\% & -24\% & -21\% & -18\% & -15\% & -7\% \\ \hline
        {\pounds}35–40k & -42\% & -40\% & -37\% & -35\% & -32\% & -28\% & -25\% & -21\% & -15\% & -7\% \\ \hline
        {\pounds}40–45k & -43\% & -42\% & -40\% & -38\% & -36\% & -33\% & -29\% & -26\% & -19\% & -10\% \\ \hline
        {\pounds}45–50k & -43\% & -42\% & -41\% & -40\% & \cellcolor{green!25} -39\% & -36\% & -33\% & -29\% & -23\% & -11\% \\ \hline
        {\pounds}50–55k & -42\% & -42\% & -41\% & -41\% &\cellcolor{green!25} -40\% & -38\% & -36\% & -33\% & -26\% & -12\% \\ \hline
        {\pounds}55–60k & -41\% & -41\% & -40\% & -40\% & -39\% & -38\% & -37\% & -34\% & -29\% & -13\% \\ \hline
        {\pounds}60–65k & -40\% & -39\% & -39\% & -39\% & -38\% & -37\% & -36\% & -34\% & -28\% & -13\% \\ \hline
        {\pounds}65–70k & -38\% & -38\% & -38\% & -37\% & -37\% & -36\% & \cellcolor{blue!25} -35\% & -33\% & -28\% & -12\% \\ \hline
        {\pounds}70–75k & -37\% & -37\% & -37\% & -36\% & -36\% & -35\% & \cellcolor{blue!25} -34\% & -32\% & -27\% & -12\% \\ \hline
        {\pounds}75–80k & -36\% & -36\% & -36\% & -35\% & -35\% & -34\% & -33\% & -31\% & -26\% & -12\% \\ \hline
        {\pounds}80–85k & -35\% & -35\% & -35\% & -34\% & -34\% & -33\% & -32\% & -30\% & -26\% & -11\% \\ \hline
        {\pounds}85–90k & -34\% & -34\% & -34\% & -33\% & -33\% & -32\% & -31\% & -30\% & -25\% & -11\% \\ \hline
        {\pounds}90–95k & -33\% & -33\% & -33\% & -33\% & -32\% & -32\% & -31\% & -29\% & -24\% & -11\% \\ \hline
        {\pounds}95–100k & -32\% & -32\% & -32\% & -32\% & -31\% & -31\% & -30\% & -28\% & -24\% & -11\% \\ \hline
        {\pounds}100–105k & -31\% & -31\% & -31\% & -31\% & -31\% & -30\% & -29\% & -28\% & -23\% & -10\% \\ \hline
        {\pounds}105–110k & -30\% & -30\% & -30\% & -30\% & -30\% & -29\% & -29\% & -27\% & -23\% & -10\% \\ \hline
        {\pounds}110–115k & -29\% & -29\% & -29\% & -29\% & -29\% & -29\% & -28\% & -27\% & -23\% & -10\% \\ \hline
        {\pounds}115–120k & -28\% & -29\% & -29\% & -29\% & -29\% & -28\% & -27\% & -26\% & -22\% & -10\% \\ \hline
        {\pounds}120–125k & -28\% & -28\% & -28\% & -28\% & -28\% & -28\% & -27\% & -26\% & -22\% & -10\% \\ \hline
        {\pounds}125–130k & -27\% & -27\% & -27\% & -27\% & -27\% & -27\% & -26\% & -25\% & -21\% & -9\% \\ \hline
        {\pounds}130–135k & -26\% & -27\% & -27\% & -27\% & -27\% & -26\% & -26\% & -25\% & -21\% & -9\% \\ \hline
        {\pounds}135–140k & -26\% & -26\% & -26\% & -26\% & -26\% & -26\% & -25\% & -24\% & -20\% & -9\% \\ \hline
        {\pounds}140–145k & -25\% & -25\% & -26\% & -26\% & -26\% & -25\% & -25\% & -24\% & -20\% & -9\% \\ \hline
        {\pounds}145–150k & -25\% & -25\% & -25\% & -25\% & -25\% & -25\% & -24\% & -23\% & -20\% & -9\% \\ \hline
        {\pounds}150+ & -20\% & -20\% & -21\% & -21\% & -21\% & -21\% & -20\% & -20\% & -17\% & -7\% \\ \hline
    \end{tabular}
    
\caption{Percentage loss for CPI 2.8\% across age and salary values of the UUK Heat map. USS's three data points, represented as personas Aria (37 yrs, {\pounds}30k), Bryn (43 yrs, {\pounds}50k), and Chloe  (51 yrs, {\pounds}70k) are shaded in respectively pink, green and blue. Data and calculations at  \href{https://github.com/SussexUCU/USS/tree/main/data/USS_modeller}{github} }
    \label{tab:percentage loss heatmap}
\end{table}

\begin{table}[!ht]
    \centering
    \begin{tabular}{|r||r|r|r|r|r|r|r|r|r|r|}
    \hline
        Salary / age & $\leq$ 25 & 25-29 & 30-34 & 35-39 & 40-44 & 45-49 & 50-54 & 55-59 & 60-64 & 65+ \\ \hline
        \hline
        {\pounds}0–5k & -{\pounds}11 & -{\pounds}9 & -{\pounds}7 & -{\pounds}6 & -{\pounds}5 & -{\pounds}3 & -{\pounds}2 & -{\pounds}1 & {\pounds}0 & {\pounds}0 \\ \hline
        {\pounds}5–10k & -{\pounds}33 & -{\pounds}27 & -{\pounds}22 & -{\pounds}17 & -{\pounds}13 & -{\pounds}9 & -{\pounds}6 & -{\pounds}4 & -{\pounds}1 & {\pounds}0 \\ \hline
        {\pounds}10–15k & -{\pounds}55 & -{\pounds}45 & -{\pounds}36 & -{\pounds}29 & -{\pounds}22 & -{\pounds}16 & -{\pounds}10 & -{\pounds}6 & -{\pounds}2 & {\pounds}0 \\ \hline
        {\pounds}15–20k & -{\pounds}77 & -{\pounds}63 & -{\pounds}51 & -{\pounds}40 & -{\pounds}30 & -{\pounds}22 & -{\pounds}15 & -{\pounds}8 & -{\pounds}3 & {\pounds}0 \\ \hline
        {\pounds}20–25k & -{\pounds}108 & -{\pounds}86 & -{\pounds}67 & -{\pounds}52 & -{\pounds}39 & -{\pounds}28 & -{\pounds}19 & -{\pounds}11 & -{\pounds}4 & -{\pounds}1 \\ \hline
        {\pounds}25–30k & -{\pounds}149 & -{\pounds}118 & -{\pounds}91 &\cellcolor{red!25} -{\pounds}68 & -{\pounds}50 & -{\pounds}35 & -{\pounds}23 & -{\pounds}13 & -{\pounds}5 & -{\pounds}1 \\ \hline
        {\pounds}30–35k & -{\pounds}194 & -{\pounds}156 & -{\pounds}122 &\cellcolor{red!25} -{\pounds}92 & -{\pounds}67 & -{\pounds}45 & -{\pounds}28 & -{\pounds}15 & -{\pounds}6 & -{\pounds}1 \\ \hline
        {\pounds}35–40k & -{\pounds}239 & -{\pounds}196 & -{\pounds}156 & -{\pounds}120 & -{\pounds}88 & -{\pounds}61 & -{\pounds}38 & -{\pounds}20 & -{\pounds}7 & -{\pounds}1 \\ \hline
        {\pounds}40–45k & -{\pounds}273 & -{\pounds}231 & -{\pounds}190 & -{\pounds}150 & -{\pounds}112 & -{\pounds}80 & -{\pounds}52 & -{\pounds}29 & -{\pounds}10 & -{\pounds}2 \\ \hline
        {\pounds}45–50k & -{\pounds}294 & -{\pounds}253 & -{\pounds}213 & -{\pounds}173 & \cellcolor{green!25}-{\pounds}135 & -{\pounds}98 & -{\pounds}65 & -{\pounds}37 & -{\pounds}13 & -{\pounds}2 \\ \hline
        {\pounds}50–55k & -{\pounds}305 & -{\pounds}265 & -{\pounds}226 & -{\pounds}187 & \cellcolor{green!25}-{\pounds}149 & -{\pounds}113 & -{\pounds}78 & -{\pounds}45 & -{\pounds}17 & -{\pounds}2 \\ \hline
        {\pounds}55–60k & -{\pounds}310 & -{\pounds}270 & -{\pounds}231 & -{\pounds}193 & -{\pounds}155 & -{\pounds}119 & -{\pounds}84 & -{\pounds}51 & -{\pounds}20 & -{\pounds}3 \\ \hline
        {\pounds}60–65k & -{\pounds}310 & -{\pounds}270 & -{\pounds}231 & -{\pounds}193 & -{\pounds}156 & -{\pounds}119 & -{\pounds}85 & -{\pounds}52 & -{\pounds}20 & -{\pounds}3 \\ \hline
        {\pounds}65–70k & -{\pounds}310 & -{\pounds}270 & -{\pounds}231 & -{\pounds}193 & -{\pounds}156 &  -{\pounds}119 & \cellcolor{blue!25}-{\pounds}85 & -{\pounds}52 & -{\pounds}21 & -{\pounds}3 \\ \hline
        {\pounds}70–75k & -{\pounds}310 & -{\pounds}270 & -{\pounds}232 & -{\pounds}193 & -{\pounds}156 &  -{\pounds}119 & \cellcolor{blue!25}-{\pounds}85 & -{\pounds}52 & -{\pounds}21 & -{\pounds}3 \\ \hline
        {\pounds}75–80k & -{\pounds}310 & -{\pounds}270 & -{\pounds}232 & -{\pounds}193 & -{\pounds}156 & -{\pounds}119 & -{\pounds}85 & -{\pounds}52 & -{\pounds}21 & -{\pounds}3 \\ \hline
        {\pounds}80–85k & -{\pounds}310 & -{\pounds}271 & -{\pounds}232 & -{\pounds}193 & -{\pounds}156 & -{\pounds}119 & -{\pounds}85 & -{\pounds}52 & -{\pounds}21 & -{\pounds}3 \\ \hline
        {\pounds}85–90k & -{\pounds}310 & -{\pounds}270 & -{\pounds}231 & -{\pounds}193 & -{\pounds}156 & -{\pounds}119 & -{\pounds}85 & -{\pounds}52 & -{\pounds}20 & -{\pounds}3 \\ \hline
        {\pounds}90–95k & -{\pounds}310 & -{\pounds}270 & -{\pounds}231 & -{\pounds}193 & -{\pounds}156 & -{\pounds}119 & -{\pounds}85 & -{\pounds}52 & -{\pounds}20 & -{\pounds}3 \\ \hline
        {\pounds}95–100k & -{\pounds}310 & -{\pounds}271 & -{\pounds}232 & -{\pounds}193 & -{\pounds}156 & -{\pounds}119 & -{\pounds}85 & -{\pounds}52 & -{\pounds}20 & -{\pounds}3 \\ \hline
        {\pounds}100–105k & -{\pounds}310 & -{\pounds}270 & -{\pounds}232 & -{\pounds}193 & -{\pounds}156 & -{\pounds}119 & -{\pounds}85 & -{\pounds}52 & -{\pounds}20 & -{\pounds}3 \\ \hline
        {\pounds}105–110k & -{\pounds}310 & -{\pounds}270 & -{\pounds}232 & -{\pounds}193 & -{\pounds}156 & -{\pounds}119 & -{\pounds}85 & -{\pounds}52 & -{\pounds}21 & -{\pounds}3 \\ \hline
        {\pounds}110–115k & -{\pounds}310 & -{\pounds}270 & -{\pounds}231 & -{\pounds}193 & -{\pounds}156 & -{\pounds}119 & -{\pounds}85 & -{\pounds}52 & -{\pounds}21 & -{\pounds}3 \\ \hline
        {\pounds}115–120k & -{\pounds}310 & -{\pounds}270 & -{\pounds}231 & -{\pounds}193 & -{\pounds}156 & -{\pounds}119 & -{\pounds}85 & -{\pounds}52 & -{\pounds}21 & -{\pounds}3 \\ \hline
        {\pounds}120–125k & -{\pounds}310 & -{\pounds}270 & -{\pounds}231 & -{\pounds}193 & -{\pounds}156 & -{\pounds}119 & -{\pounds}85 & -{\pounds}52 & -{\pounds}21 & -{\pounds}3 \\ \hline
        {\pounds}125–130k & -{\pounds}310 & -{\pounds}270 & -{\pounds}232 & -{\pounds}193 & -{\pounds}156 & -{\pounds}119 & -{\pounds}85 & -{\pounds}52 & -{\pounds}21 & -{\pounds}3 \\ \hline
        {\pounds}130–135k & -{\pounds}310 & -{\pounds}271 & -{\pounds}232 & -{\pounds}193 & -{\pounds}156 & -{\pounds}119 & -{\pounds}85 & -{\pounds}52 & -{\pounds}20 & -{\pounds}3 \\ \hline
        {\pounds}135–140k & -{\pounds}310 & -{\pounds}270 & -{\pounds}232 & -{\pounds}193 & -{\pounds}156 & -{\pounds}119 & -{\pounds}85 & -{\pounds}52 & -{\pounds}20 & -{\pounds}3 \\ \hline
        {\pounds}140–145k & -{\pounds}310 & -{\pounds}270 & -{\pounds}231 & -{\pounds}193 & -{\pounds}156 & -{\pounds}119 & -{\pounds}85 & -{\pounds}52 & -{\pounds}20 & -{\pounds}3 \\ \hline
        {\pounds}145–150k & -{\pounds}310 & -{\pounds}271 & -{\pounds}231 & -{\pounds}193 & -{\pounds}156 & -{\pounds}119 & -{\pounds}85 & -{\pounds}52 & -{\pounds}20 & -{\pounds}3 \\ \hline
        {\pounds}150+ & -{\pounds}310 & -{\pounds}271 & -{\pounds}232 & -{\pounds}193 & -{\pounds}156 & -{\pounds}119 & -{\pounds}85 & -{\pounds}52 & -{\pounds}20 & -{\pounds}3 \\ \hline
    \end{tabular}
\caption{Loss in thousands of pounds in today's money for CPI 2.8\% across age and salary values of UUK Heat map. USS's three data points, represented as personas Aria (37 yrs, {\pounds}30k), Bryn (43 yrs, {\pounds}50k), and Chloe  (51 yrs, {\pounds}70k) are shaded in respectively pink, green and blue. Data and calculations at \href{https://github.com/SussexUCU/USS/tree/main/data/USS_modeller}{github} }
    \label{tab:pound loss heatmap}

\end{table}

\end{document}